\documentclass[12pt]{iopart}
\usepackage[utf8]{inputenc}

\usepackage[
backend=bibtex,
style=numeric,
sorting=none
]{biblatex}
\usepackage{graphicx}
\expandafter\let\csname equation*\endcsname\relax
\expandafter\let\csname endequation*\endcsname\relax

\usepackage{amsmath}
\usepackage{amssymb}
\usepackage{hyperref}
\usepackage{mathrsfs}

\graphicspath{{figures/}}

\newcommand{\psih}{\hat{\psi}}
\newcommand{\psihd}{\hat{\psi}^\dagger}

\DeclareFontEncoding{LS1}{}{}
\DeclareFontSubstitution{LS1}{stix}{m}{n}
\DeclareSymbolFont{symbols2}{LS1}{stixfrak}{m}{n}

\DeclareMathSymbol{\lcurvyangle}{\mathopen}{symbols2}{"E9}
\DeclareMathSymbol{\rcurvyangle}{\mathclose}{symbols2}{"EA}

\newcommand{\dlangle}{\lcurvyangle}
\newcommand{\drangle}{\rcurvyangle}

\newcommand{\Sigmacal}{\mathit{\Sigma}}

\renewcommand{\S}{\Sigmacal}

\newcommand{\Ical}{\mathcal{I}}

\newcommand{\Mcal}{{\mathcal{M}}}

\newcommand{\Ah}{\hat{A}}
\newcommand{\Uh}{\hat{U}}

\newcommand{\Hh}{\hat{H}}

\newcommand{\rhoh}{\hat{\rho}}

\newcommand{\Nh}{\hat{N}}

\newcommand{\trace}[1]{\Tr\left [ #1 \right ]}

\newcommand{\Eq}[1]{\eqref{#1}}

\newcommand{\xb}{\mathbf{x}}

\newcommand{\la}{\langle}
\newcommand{\ra}{\rangle}

\newcommand{\cop}[2]{\hat{#1}^\dagger_{#2}}

\newcommand{\aop}[2]{\hat{#1}_{#2}}

\newcommand{\bigdiagram}[1]{\includegraphics[scale=1.0]{figures/#1.pdf}}

\newcommand{\bigdiagramcenter}[1]{\vcenter{\hbox{\bigdiagram{#1}}}}

\newcommand{\Ncal}{\mathcal{N}}

\newcommand{\Lcal}{\mathcal{L}}

\newcommand{\bfx}{\mathbf{x}}
\newcommand{\bfr}{\mathbf{r}}

\newcommand{\Bh}{\hat{B}}
\newcommand{\Fh}{\hat{F}}

\newcommand{\irr}{{\mathrm{irr}}}

\begin{document}

\title{Cutting rules and positivity in finite temperature many-body theory}

\author{M. J. Hyrk{\"a}s, D. Karlsson, and R. van Leeuwen}
\address{Department of Physics,
Nanoscience Center P.O.Box 35 FI-40014 University of Jyvaskyla, Finland}
\ead{markku.j.hyrkas@jyu.fi}

\begin{abstract}
For a given diagrammatic approximation in many-body perturbation theory it is not guaranteed that positive observables, such as the density or the spectral function, retain their positivity. 
For zero-temperature systems we developed a method [Phys.Rev.B{\bf 90},115134 (2014)] based on so-called cutting rules for Feynman diagrams that enforces these properties diagrammatically, thus solving the problem of negative spectral densities observed for various vertex approximations.
In this work we extend this method to systems at finite temperature by formulating the cutting rules in terms of retarded $N$-point functions,
thereby simplifying earlier approaches and simultaneously solving the issue of non-vanishing vacuum diagrams that has plagued finite temperature expansions. 
Our approach is moreover valid for nonequilibrium systems in initial equilibrium and allows us to show that important commonly used approximations, namely the $GW$, second Born and $T$-matrix approximation, retain positive spectral functions at finite temperature.
Finally we derive an analytic continuation relation between the spectral forms of retarded $N$-point functions and their Matsubara counterparts and a set of Feynman rules to evaluate them.  

\end{abstract}

\noindent{\it Keywords\/}: diagrammatic perturbation theory, non-equilibrium Green’s functions, quantum many-body theory, spectral properties

\submitto{\jpa}
\maketitle

\section{Introduction}\label{sec:1_introduction}

Non-equilibrium Green's function theory \cite{Stefanucci2013} is a powerful tool for calculating time-dependent properties in a variety of quantum many-particle systems. While exact in principle, in practice the formalism relies on approximate diagrammatic expansions; to finite order, or to infinite order via resummations.

For a given diagrammatic approximation, there is in general no guarantee that relevant properties of the exact solution will be retained. This, in particular, applies to the positivity of probability distributions, such as the spectral function in energy space and the particle density in position space.
For example when the lowest order vertex correction in a dynamically screened expansion for the electron gas was considered, the spectral function was found to become negative~\cite{Minnhagen1974} with a similar issue occurring for the absorption spectrum~\cite{Brosens1984}. The same problems were observed also in finite systems for the case of atoms and small Hubbard lattices~\cite{Hellgren2009,Schindlmayr1998}. 
Injudiciously chosen approximations may even result in both the spectral function and the particle density becoming negative, as was demonstrated for the Anderson model~\cite{Karlsson2016c}. These issues have considerably hindered progress in the study of spectral properties beyond the simplest approximations.

For the case of equilibrium systems at zero temperature the problem of negative spectral densities was solved with
a systematic diagrammatic method~\cite{Stefanucci2014,Uimonen2015} in which Feynman graphs are expressed in terms of so-called half-diagram products that could be derived from the Lehmann representation of the correlation functions. Diagrammatically 
this procedure amounts to cutting the Feynman graphs in various ways and gluing the pieces together in order to form new manifestly positive products such that
a non-negative spectral function is guaranteed. 
Of practical importance is the possibility \cite{Stefanucci2014} to extend a given non-positive approximation with a minimal set of extra diagrams to enforce the positivity condition. This constitutes the so-called Positive Semi-Definite (PSD) expansion for spectral functions.
Another considerable advantage of the method is that it allows for an expansion in physical scattering processes, which was successfully used to study the various contributions of particle-hole and plasmon excitations to the spectral function of the electron gas \cite{Pavlyukh2016,Pavlyukh2020}.
A similar technique was also developed for the steady state limit of non-equilibrium systems initially in a zero-temperature equilibrium~\cite{Hyrkas2019b}.
This extension is not straightforward as it required a new expansion technique in terms of so-called multi-retarded half-diagrams and appears as a special case of a general integral calculus for multi-argument contour functions, originally developed by Danielewicz~\cite{Danielewicz1990} and extended by us~\cite{Hyrkas2019}.

The previously mentioned developments leave out the important case of systems in finite temperature equilibrium, which will form the main topic of this paper.
Finite temperature many-body theory is vital for the description of excited state properties in warm dense matter systems \cite{Kas2017}, such as laser-shocked, fusion and astrophysical systems, and phase transitions in nuclear matter \cite{Rios2008}. It is
also important to describe coupled electron-boson systems, for example in the study of temperature-dependent properties of solid state systems in which electron-phonon interactions are crucial \cite{Giustino2017,pavlyukh2021,Karlsson2021}. Another relevant class of physical systems in this respect is that of small electronic systems coupled to baths \cite{Zgid2017}, such as nanojunctions \cite{Galperin2009}.

In the particle physics community several diagrammatic cutting approaches have been developed to study positive scattering amplitudes in finite temperature quantum field theory but they turn out to have undesirable features that make them unsuitable for our purpose of constructing a PSD perturbation theory. The method we develop in this paper connects and simplifies some of these approaches and it is therefore useful to give a brief overview of them. 

Early works by Kobes and Semenoff \cite{Kobes1985,Kobes1986} (see also~\cite{Gelis1997} for an overview) are based 
on the largest-time equation originally
invented by Veltman \cite{Veltman1963,Veltman1994} and derive expansions in time-ordered and anti-time-ordered diagrams, containing so-called "non-cuttable" diagrams \cite{Kobes1986} which generate disconnected subdiagrams when the standard cutting rules are applied. This, in turn, leads to the appearance of disconnected vacuum diagrams in the construction
of PSD approximations which is undesirable as they are absent in an exact expansion. Later work \cite{Bedaque1997,Gelis1997} based on the same method managed to rearrange diagrams in such a way that non-cuttable pieces do not arise at the expense of a proliferation of extra Green's function lines making the method very laborious and equally unsuitable for our purpose of creating a finite temperature PSD perturbation theory.
A very different approach was pursued by Jeon \cite{Jeon1993} based solely on Matsubara diagrams, but leading eventually to an expansion in (anti)-time ordered diagrams again with the appearance of disconnected diagrams in the same vein as earlier works. 
Finally Landshoff \cite{Landshoff1996}, with the example of bosonic particles, gave a simpler derivation purely based on Keldysh contour integrals and the Lehmann representation but also ends up with an expansion involving disconnected subdiagrams after cutting. Our earlier approach for zero-temperature systems is closely related to this derivation but in the present work we use a different and simpler procedure to generalize our earlier method \cite{Stefanucci2014} to the case of finite temperature systems. Furthermore, to solve the critical issue of the non-cuttable diagrams we use the technique of retarded half-diagrams \cite{Hyrkas2019,Hyrkas2019b} instead of the (anti)-time ordered expansion of Landshoff.

In hindsight it is found that all earlier approaches mentioned above can be obtained in a rather simple way
from the more straightforward derivation which forms the main subject of this paper.
One of our goals therefore is to put earlier work into a new context and to elucidate its relations to the present work.
As it turns out, our approach can also be naturally generalized to the case of finite temperature systems at initial equilibrium which are subsequently perturbed into a non-equilibrium state. We can for this general case demonstrate the positivity of spectral functions for several commonly used approximations, i.e. the GW, second Born and T-matrix approximations.

The paper is outlined as follows. In Chapter 2 we discuss the structure of exact correlators and their relation to the positivity of spectral functions. 
In Chapter 3 we
derive a PSD perturbation theory for the self-energy of finite temperature systems 
in terms of retarded half-diagram products, thereby generalizing earlier work for zero-temperature systems~\cite{Stefanucci2014,Hyrkas2019}. Finally we apply the new formalism in Chapter 4 to demonstrate the positivity of the GW, second Born and the T-matrix approximations for finite temperature systems, before concluding in Chapter 5.

\section{Positivity in many-body theory}

\subsection{General framework}

\label{sec:inner_product_form}

We start with a brief introduction to the background theory and consider a system of fermions interacting via a two-body interaction described by the Hamiltonian
\begin{equation} \begin{split} \label{eq:hamiltonian}
\hat{H}(t) = \int \rmd \xb \ \psihd(\xb) h(\xb,t) \psih(\xb)
+
\frac{1}{2}\int \rmd \xb \rmd \xb' \ \psihd(\xb) \psihd(\xb') v(\xb,\xb') \psih(\xb') \psih(\xb),
\end{split} \end{equation}
where $h(\xb,t)$ is an unspecified time-dependent one-body operator, and $v(\xb,\xb')$ a general two-body interaction. The field operators  $\psihd(\xb)$ and $\psih(\xb)$ respectively create and annihilate a particle at space-spin point $\xb = (\bfr,\sigma)$. 
For an equilibrium system at finite temperature we can specify an initial ensemble described by a density operator of the form 
\begin{equation}
\rhoh = \frac{\rme^{-\beta \hat{H}_M }}{ Z}
\label{exponential_rho}
\end{equation}
with the partition function $Z=\trace{\rme^{-\beta \hat{H}_M }}$ and the Matsubara Hamiltonian
\begin{equation}
\hat{H}_M = \hat{H} (t_0) - \mu \hat{N},
\label{H_Matsubara}
\end{equation}
where $\hat{H} (t_0)$ is the Hamiltonian of \Eq{eq:hamiltonian} evaluated at an initial time $t_0$, $\hat{N}$ is the number operator, and $\mu$ the chemical potential. A strictly positive density operator, i.e. satisfying
$\langle \varphi | \hat{\rho} |\varphi \rangle >0$ for any non-zero state $| \varphi \rangle$, can always be written in the form of \Eq{exponential_rho} albeit that $\hat{H}_M$ will then in general be an $N$-body operator rather than the simpler $2$-body operator of \Eq{H_Matsubara} \cite{Stefanucci2013,vanLeeuwen2012,vanLeeuwen2013,Garny2009,Wagner1991}. 

\begin{figure*}
\centering
\includegraphics[scale=1.4]{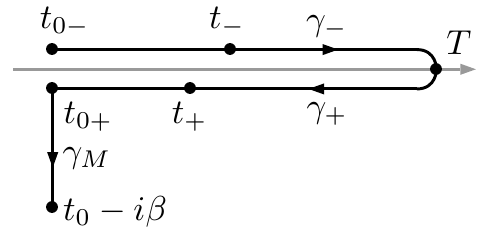}
\caption{The contour $\gamma$, with a forward ($\gamma_-$) and backward ($\gamma_+$) real-time branches (drawn off the real axis to distinguish them) and an imaginary Matsubara branch ($\gamma_M$). Times on the forward and backward branches are denoted by $t_-$ and $t_+$ respectively. \label{fig:contour}} 
\end{figure*}

If $\hat{O} (t)$ is an (in general time-dependent) operator in the Schr\"odinger picture, then its time-dependent expectation value is given by
\begin{equation}
\langle O \rangle (t) = \trace{ \rhoh \, \hat{U} (t_0,t) \hat{O} (t) \hat{U} (t,t_0)} = \trace{ \rhoh \, \hat{O}_H (t) } 
\end{equation}
where $\hat{U} (t,t_0)$ is the time-evolution operator that evolves the system from time $t_0$ to time $t$; its explicit form involves a time-ordered exponential where for details we refer to the literature \cite{Stefanucci2013}.
For carrying out perturbative expansions it is advantageous to re-express
the expectation value in a contour ordered form as
\begin{equation}
    \langle O \rangle (z) = \trace{ \mathcal{T}_\gamma \{ \rme^{-\rmi \int_\gamma d\bar{z} \hat{H} (\bar{z})} \hat{O} (z) \} } 
\end{equation}
where $\gamma$ is the time-contour in Figure \ref{fig:contour}
consisting of a forward branch $\gamma_-$ and a backward branch $\gamma_+$ in the time interval $[t_0,T]$
and a vertical or Matsubara track in the complex time interval $[t_0,t_0-i\beta]$ on which the Hamiltonian is given by $\hat{H}_M$. The time $T$ is any time after which the expectation value is to be evaluated and often is taken to be infinity for convenience \cite{Stefanucci2013}.
Perturbative expressions are then derived from this formula by expanding the time-ordered exponential in powers of the two-body interaction.

The basic ingredients of diagrammatic perturbation theory are contour ordered products of operators defined on $\gamma$, the simplest ones being strings of just two operators $\hat{O}_1$ and $\hat{O}_2$ of the general form
\begin{equation}
   k(z,z') =  \trace{ \rhoh \, \mathcal{T}_\gamma \{ \hat{O}_1 (z) \hat{O}_2 (z') \} } = \theta (z,z') k^> (t,t') + \theta (z',z) k^< (t,t')
\end{equation}
where $\theta (z,z')$ is a contour Heaviside function equal to one
if $z$ is later than $z'$ (symbolically denoted  by $z > z'$)
in contour ordering and zero otherwise \cite{Stefanucci2013}. A contour time $z$ corresponding to real time $t$ is denoted by $t_-$ if it occurs
on $\gamma_-$ and $t_+$ if it occurs on $\gamma_+$. All operators
that we consider satisfy $\hat{O}(t_\pm)=\hat{O}(t)$
i.e. they assume the same value on both horizontal branches. Consequently
the functions $k^\lessgtr$ are real time functions of the explicit form
\begin{equation}
    k^>(t,t') = 
    \trace{ \rhoh \,  \hat{O}_1 (t) \hat{O}_2 (t') }
    \quad \quad 
    k^<(t,t') = 
    \pm \trace{ \rhoh \,  \hat{O}_2 (t') \hat{O}_1 (t) }
\end{equation}
($+/-$ for bosonic/fermionic operators) which we will refer to as many-body correlation functions, or simply correlators.
An important case is when $\hat{O}_1 (z)=\psih_H (\xb,z)$ and
$\hat{O}_2 (z)=\psihd_H (\xb',z)$ in which case the functions
$k^>$ and $k^<$ correspond to the functions $\rmi G^>$ and $\rmi G^<$
representing the particle and hole Green's functions. The way these
correlation functions incorporate positivity constraints on physical observables is discussed in the next section.

\subsection{Positivity constraints on exact correlators}
\label{sec:propExactSolution}

In this section we expose a general structure of non-equilibrium correlators, namely that of a positive semi-definite Hermitian form, and show that this structure is sufficient to guarantee important positivity constraints. By casting approximate diagrammatic theories in this form these properties are then automatically satisfied in the approximate theory. We delay the discussion of approximations to the following chapters and will first discuss the case of exact correlators. Let $\hat{\rho}$ be a density operator
of the form
\begin{equation}
\hat{\rho} = \sum_j w_j | \Psi_j \rangle \langle  \Psi_j |,
\end{equation}
where the occupation numbers $w_j \geq 0$, subject to the condition $ \sum_j w_j =1$, describe a probability distribution over a set $\left\{  | \Psi_j \rangle \right\}$ 
of normalized initial many-body states. This defines a positive semi-definite (PSD) operator $\rhoh$ by which we mean that 
\begin{equation}
\langle \varphi | \hat{\rho} | \varphi \rangle \geq 0
\label{operatorPSD}
\end{equation}
for any state $| \varphi \rangle$ in the Hilbert space.
We further consider two general operators $\hat{A}$ and $\hat{B}$ in Hilbert space. These operators may depend on various parameters, such as space and time, but these will be suppressed as they are not relevant at this point of the discussion.
We define the (weighted) Hilbert-Schmidt product \cite{Haag1967} of these operators as:
\begin{equation}
 \dlangle \Ah | \Bh \drangle = \trace{\rhoh \Ah^\dagger \Bh},
\end{equation}
where curvy angled brackets are used to distinguish the Hilbert-Schmidt product from the standard Hilbert space inner product.
The Hilbert-Schmidt product is well-defined if $\hat{A}$ and $\hat{B}$ are bounded operators, as is discussed in detail by Haag, Hugenholtz and Winnink \cite{Haag1967}.
As demonstrated in \ref{app:PSDHF} this product satisfies the properties of a positive semi-definite Hermitian form (PSDHF) when the operator $\hat{\rho}$ is positive semi-definite, and that of an inner product when $\hat{\rho}$ is strictly positive. For the purposes of this work the positive semi-definiteness is both sufficient and most practical as we will often encounter finite diagrammatic expansions that represent a PSDHF but do not possess the strict positivity property.
A consequence of the PSDHF structure is that the Cauchy-Schwartz inequality
\begin{equation}
 |\dlangle \Ah | \Bh \drangle|^2
 \leq
 \dlangle \Ah | \Ah \drangle \dlangle \Bh | \Bh \drangle
\end{equation}
is satisfied (see \ref{app:PSDHF} for a short proof) which leads to useful constraints on the correlators.
The positive semi-definiteness of the Hilbert-Schmidt product furthermore leads to interesting corollaries by choosing particular forms of the operator $\Ah$. 
If we take 
\begin{equation}
\hat{A}= \int \rmd\xb \, \varphi (\xb) \, \psih_H (\xb, t)
\end{equation}
we obtain
\begin{align}
    0 \leq \dlangle \hat{A} | \hat{A} \drangle 
    = \int \rmd\xb \rmd\xb' \, \varphi^* (\xb)
    \trace{ \rhoh \, \psihd_H (\xb,t) \psih_H (\xb',t)}
    \varphi (\xb')
\end{align}
which expresses the fact that at each time $t$ the one-particle density matrix, regarded as an integral kernel acting on spatial functions, is a positive semi-definite operator.
This is an important property of the density matrix. It guarantees, for example, that the instantaneous natural orbital occupation numbers \cite{Appel2010} 
obtained by diagonalizing the density matrix are non-negative.
Another important case arises if we take
\begin{equation}
    \hat{A} = \int \rmd t \, \varphi (t) \hat{O}_H (t)
\end{equation}
 where $\hat{O}_H (t)$ is the Heisenberg form of a Schr\"odinger operator $\hat{O}$.
In that case we obtain
\begin{equation}
    0 \leq \dlangle \hat{A} | \hat{A} \drangle =
    \int \rmd t \rmd t'\, \varphi^* (t) C(t,t') \varphi (t')
    = \int \frac{\rmd\omega}{2\pi} \frac{\rmd\omega'}{2\pi} \tilde{\varphi}^* (\omega) \tilde{C} (\omega,\omega') \tilde{\varphi} (\omega') 
\label{ft_psd}
\end{equation}
where $\tilde{\varphi}(\omega)$ is the Fourier transform of $\varphi$ and we defined
\begin{equation}
C(t,t') = \dlangle \hat{O}_H (t) | \hat{O}_H (t') \drangle = \trace{\hat{\rho} \, \hat{O}_H^\dagger (t) \hat{O}_H (t') }
\label{Ctt}
\end{equation}
as well as its Fourier transform
\begin{equation}
 \tilde{C} (\omega,\omega') = \int \rmd t \rmd t' \, \rme^{\rmi (\omega t - \omega' t')} \, C(t,t').
 \label{C_ft}
\end{equation}
Let us now investigate the equilibrium or steady-state limit. It is convenient to introduce the relative time-coordinate $\tau = t- t'$
and the average time coordinate $T=(t+t')/2$, which transforms the expression in \Eq{C_ft} to
\begin{equation}
 \tilde{C} (\omega,\omega') = \int \rmd\tau \rmd T \, \rme^{\rmi (\omega - \omega' ) T + \rmi(\omega+\omega') \tau /2} \, C(T + \frac{\tau}{2}, T - \frac{\tau}{2}).
 \label{C_ft2}
\end{equation}
For equilibrium systems or in the steady state limit of a non-equilibrium system \cite{Hyrkas2019b} the correlator is independent of $T$ and the equation becomes
\begin{equation}
 \tilde{C} (\omega,\omega') = 2 \pi \delta (\omega - \omega')  \, \mathcal{A} (\omega)
  \label{C_ft3}
\end{equation}
where we defined
\begin{equation}
\mathcal{A} (\omega) = \int \rmd\tau\, \rme^{  \rmi\omega \tau } \, C(\tau, 0) =  \int \rmd\tau\, \rme^{  \rmi\omega \tau } \,  \dlangle \hat{O}_H (\tau) | \hat{O}_H (0) \drangle.
\label{gen_spec}
\end{equation}
For a general correlator of the form of \Eq{Ctt} in an equilibrium or steady state regime the function $\mathcal{A} (\omega)$ defined
by \Eq{gen_spec} is called the corresponding {\em spectral function}.
If we now insert expression \Eq{C_ft3} back into \Eq{ft_psd} we find that
\begin{equation}
\int \frac{\rmd\omega}{2\pi} \,  | \tilde{\varphi} (\omega) |^2 \, \mathcal{A} (\omega) \geq 0.
\label{ft_psd2}
\end{equation}
Since this expression is valid for a general function $\tilde{\varphi}$ it follows that the spectral function is pointwise (for each given $\omega$) PSD, i.e.
\begin{equation}
 \mathcal{A} (\omega) \geq 0,
\label{a_psd}
\end{equation}
which is an important consequence of the Hilbert-Schmidt product structure of the correlator in \Eq{Ctt}.
As an illustration we examine the spectral function corresponding to $\rmi G^>$. To this end we consider \Eq{gen_spec} and take the operator $\hat{O}$ 
in this expression to be of the form
\begin{equation}
\hat{O} = \int \rmd\xb \, u(\xb) \, \psihd (\xb)
\end{equation}
with $u (\xb)$ a general function. Equations \Eq{gen_spec} and 
\Eq{a_psd} then yield
\begin{equation}
\mathcal{A} (\omega) =  \int \rmd\xb \rmd\xb' \, u^* (\xb) A^> (\xb,\xb'; \omega) u(\xb') \geq 0
\label{spec_gt}
\end{equation}
where we defined the matrix spectral function
\begin{align}
A^> (\xb,\xb';\omega) &=  \int \rmd\tau\, \rme^{  \rmi\omega \tau } \,  \dlangle \psihd_H (\xb,\tau) | \psihd_H (\xb',0) \drangle  \nonumber \\
&= \rmi  \int \rmd\tau\, \rme^{  \rmi\omega \tau } \, G^>(\xb \tau, \xb' 0) = \rmi G^> (\xb,\xb'; \omega)
\end{align}
in which on the last line we also defined the Fourier transform of the particle propagator.
Since the expression in \Eq{spec_gt} is equal to its complex conjugate it follows that $A^>$ is a Hermitian integral kernel acting in real space.
Similarly we can define a spectral function $A^<$ for the hole propagator as
\begin{align}
A^< (\xb,\xb';\omega) =  -\rmi G^< (\xb,\xb'; \omega)
=   \int d\tau\, \rme^{ \rmi\omega \tau } \,
\dlangle \psih_H (\xb',0) | \psih_H (\xb,\tau) \drangle
\end{align}
which again is a PSD operator in the same sense as in \Eq{spec_gt}.
%
In the following chapters we will simply regard the functions $G^\lessgtr$ as matrices and suppress the spatial labels $\xb$ and $\xb'$ or any other non-temporal basis labels, reintroducing them when necessary.
The particular combination
\begin{equation} \label{eq:spectral_function}
A(\omega) = \rmi \left( G^>(\omega) - G^<(\omega) \right)
\end{equation}
combines the spectral functions of the particle and hole propagator and is commonly referred to as the spectral function corresponding to
the (retarded) Green's function. This function encodes information on the
probability of many-body scattering processes as occurring in photo-emission and inverse photo-emission experiments \cite{Reinert2005, almbladh1982, Almbladh2006} and its PSD structure therefore guarantees that the probabilities are always positive semi-definite.

The PSD Hermitian form structure is very general as it applies to both equilibrium and non-equilibrium systems. In the following chapters we will investigate how this structure can be build into diagrammatic perturbation theory and thereby enforce the required positivity properties.

\subsection{Construction of PSD approximations}

In this section we briefly review the approach that we used in our earlier works to construct PSD approximations for systems at zero temperature. In practice, the spectral functions for the Green's function are not calculated directly but from a diagrammatic approximation to the self-energy appearing in the Dyson equation \cite{Stefanucci2013} written in matrix form as
\begin{equation}
G (z_1,z_2)= g(z_1,z_2) + \int_\gamma \rmd z_3 \rmd z_4 \, g(z_1,z_3) \Sigma (z_3,z_4) G(z_4,z_2)
\end{equation}
where $\Sigma$ denotes the self-energy, describing the irreducible particle scatterings, and $g$ is the Green's function of the system in the absence of interactions.
For equilibrium systems at zero temperature there is a simple relation between the greater and lesser Green's functions and the corresponding components of the correlation part of the self-energy, which in frequency space in matrix notation reads
\begin{equation}
    G^\lessgtr (\omega) = G^R (\omega) \Sigma_c^\lessgtr (\omega) G^A (\omega).
    \label{eq:Glessgtr}
\end{equation}
Here $G^R$ and $G^A$ are the retarded and advanced Green's functions which are each others adjoint. Hence to ensure the PSD property of $\rmi G^> (\omega)$  and $-\rmi G^< (\omega)$ it is sufficient
to establish the positivity of $\rmi \Sigma_c^> (\omega)$  and $-\rmi \Sigma_c^< (\omega)$.
We showed \cite{Stefanucci2014} that this can be done in a diagrammatic fashion and demonstrated, for example, that the exact $\Sigma_c$ can be written in a Hermitian product form
in which each factor can be identified with a diagrammatic
expression belonging to a cut self-energy diagram, a so-called half-diagram, a procedure that we will briefly summarize in the next section.
More importantly, we further established that, if a given approximation 
for the self-energy does not have this structure, the diagrammatic series can be extended by construction of Hermitian products to form a new approximation that is PSD.

The central open question now is whether this procedure can be extended to finite temperature systems. 
For this we will need cutting rules that can be applied to diagrams on the extended contour and which allow for the derivation of a PSD expression. In the next chapter we will derive such rules.

\section{Self-energy cutting rules at finite temperature}

The purpose of this chapter is threefold. First, we generalize our original derivation of cutting rules \cite{Stefanucci2014}, based on the Lehmann representation and the adiabatic assumption, from zero temperature to finite temperature. This is insightful since it shows that the finite temperature correction for the Lehmann amplitudes can be interpreted as an additional interaction with heat bath particles. 
Second, in the subsequent section we present an alternative derivation that uses neither the Lehmann representation nor
requires the adiabatic assumption and moreover is valid for non-equilibrium final states. 
Both derivations lead to the same expansion in time-ordered and anti-time-ordered Green's functions.
However, unlike the zero-temperature case we find that at finite temperature
it is in general difficult to derive approximate positive definite expressions without introducing unwanted vacuum diagrams in the expansion.
Third, we then demonstrate how this issue can be resolved using an expansion in so-called retarded half-diagrams \cite{Hyrkas2019,Hyrkas2019b} with a clear physical interpretation as collective contributions of past scattering processes.   

\subsection{Cutting rules from the Lehmann representation}
\label{sec:cutting_lehmann}

The correlation self-energy at finite temperature can be written as (see \cite{Danielewicz1984} or \cite{Stefanucci2013} section 9.1.)
\begin{equation} \label{eq:correlation_self_energy}
\Sigma_c(\bfx_1z_1,\bfx_2z_2) = -\rmi \textrm{Tr} \left[ \rhoh\, \mathcal{T}_\gamma \left\{ \aop{\gamma}{H}(\bfx_1z_1) \cop{\gamma}{H}(\bfx_2z_2) \right\} \right]_{\irr},
\end{equation}
where the sub-index $\irr$ denotes an operation that removes all reducible diagrams, i.e. those in which the external vertices $1$ and $2$ (corresponding to $\xb_1z_1$ and $\xb_2z_2$ respectively) can be disconnected from each other by removing a single $g$-line. The contour $\gamma$ consists of forward and backward real-time branches (see Figure \ref{fig:contour}) where in this case we do not introduce the Matsubara branch but instead work directly with the density matrix $\rhoh$. The operator $\aop{\gamma}{H}$ is the Heisenberg form of the operator
\begin{equation} \label{eq:gamma}
\hat{\gamma} (\bfx) = \int \rmd\bfx' \, v(\bfx,\bfx^\prime) \hat{\psi}^\dagger (\bfx') \hat{\psi} (\bfx^\prime) \hat{\psi} (\bfx)
= \sum_{j_1 j_2 j_3} \gamma_{j_1 j_2 j_3} (\bfx )\,  \hat{c}^\dagger_{j_1} \hat{c}_{j_2} \hat{c}_{j_3} ,
\end{equation}
where the second equality expresses the operator in a single particle basis $\varphi_j (\bfx)$ where
\begin{equation}
    \gamma_{j_1 j_2 j_3} (\bfx ) = \int \rmd\bfx' v(\bfx,\bfx') \varphi_{j_1}^*(\bfx')
    \varphi_{j_2}(\bfx') \varphi_{j_3}(\bfx)
\end{equation}
and $\hat{c}_j$ and $\hat{c}_j^\dagger$ are the creation and annihilation operators in one-particle basis.
To establish a connection to our earlier work we start with a short description of the zero-temperature
case and subsequently point out the problems that arise when attempting to generalise to the finite temperature case.

In the zero-temperature limit the density matrix $\hat{\rho}$ reduces to
$\hat{\rho} = | \Psi_0 \rangle \langle \Psi_0 |$ where $|\Psi_0 \rangle$ is the many-body ground state.
Taking the lesser component of the self-energy we can then write
\begin{equation}
    -\rmi \Sigma_c^< (1,2) = \langle \Psi_0 | \hat{\gamma}^\dagger_H (2) 
    \hat{\gamma}_H (1) | \Psi_0 \rangle_{\irr}
\end{equation}
where we used the short-hand notation $1=\bfx_1 t_1$ and $2=\bfx_2 t_2$.
Following \cite{Stefanucci2014} we first use the Gell-Mann-Low theorem \cite{Gell-Mann1951} and a time-evolution operator $\hat{U}$ to connect $| \Psi_0 \rangle$ adiabatically to a non-interacting ground state $| \Phi_0 \rangle$ at time $-T$,
 i.e. $| \Psi_0 \rangle = \hat{U}_\eta (t_0, -T) | \Phi_0 \rangle$, where eventually we let $T$
approach infinity and $\eta$ is an adiabatic parameter. This allows us to write
\begin{align}
    - \rmi \Sigma_c^< (1,2) &= \langle \Phi_0 | 
    \hat{U} (-T, t_2) \hat{\gamma}^\dagger (\bfx_2) \hat{U} (t_2, T) \hat{U} (T,t_1) \hat{\gamma} (\bfx_1)
    \hat{U} (t_1,-T) | \Phi_0 \rangle_{\irr}
    \label{eq:sigless1}
\end{align}
where we have split the evolution operator $\hat{U} (t_2,t_1)=\hat{U} (t_2, T) \hat{U} (T,t_1)$ between the
$\hat{\gamma}$ and $\hat{\gamma}^\dagger$ operators.
Then we consider a complete set of non-interacting many-body eigenstates $|L,N \rangle$ of the form
\begin{equation}
    | L, N \rangle = \hat{c}_{i_1^\prime} \ldots \hat{c}_{i_{N+1}^\prime} \hat{c}^\dagger_{i_N} \ldots 
    \hat{c}^\dagger_{i_1} | \Phi_0 \rangle
\end{equation}
where $L=(I,I')$ is a multi-index with $I=(i_1,\ldots,i_N)$ and $I'=(i^\prime_1,\ldots,i^\prime_{N+1})$. We only consider states that contain one more removed particle than an added one, as  due to the specific form of $\hat{\gamma}$ only such states
give a non-vanishing contribution when we insert them later using a completeness relation.
The states $| L, N \rangle$ satisfy the orthonormality relations
\begin{equation}
    \langle L_1, N_1 | L_2, N_2 \rangle = \delta_{N_1,N_2} \sum_P (-1)^{|P|} \delta_{L_1, P(L_2)} 
\end{equation}
where $P(L)=(P_1(I),P_2(I'))$ consists of all permutations $P_1$ and $P_2$ of the labels $I$ and $I'$ in $L$ separately and $|P|=|P_1|+|P_2|$ is the overall sign of the permutation.
There are $N! (N+1)!$ of such permutations and the completeness relation
in the relevant Hilbert space for our states is therefore given by
\begin{equation}
    \sum_{N=0}^\infty \sum_L \frac{1}{N! (N+1)!} | L, N \rangle \langle L, N | = 1
    \label{eq:completeness}
\end{equation}
where the summation index $L=(I,I')$ runs over all orderings of the indices in the multi-labels $I$ and $I'$. We now insert the completeness relation (\ref{eq:completeness}) in between the operators
$\hat{U} (t_2,T)$ and $\hat{U} (T,t_1)$ in \Eq{eq:sigless1} to obtain the expression 
\begin{align} \label{eq:sigless2}
    - \rmi \Sigma_c^< (1,2) &=
    \sum_{N=1}^\infty \sum_L \frac{1}{N! (N+1)!}  A_{L,N} (1) B_{L,N} (2)
\end{align}
which is in so-called Lehmann representation form and where we defined
\begin{align}
    &A_{L,N} (1) = \langle L,N | \hat{U} (T,t_1) \hat{\gamma} (\bfx_1)
    \hat{U} (t_1,-T) | \Phi_0 \rangle_{\irr}  \label{eq:A_def} \\
    &B_{L,N} (2) = \langle \Phi_0 | \hat{U} (-T,t_2) \hat{\gamma}^\dagger (\bfx_2)
    \hat{U} (t_2,T) | L,N \rangle_{\irr} \label{eq:B_def}.
\end{align}
These functions are related by complex conjugation,
i.e $A_{L,N}(1)=B_{L,N}^* (1)$,
but we give them different names since they have different diagrammatic expansions; the function $A_{L,N}$ can be expanded in time-ordered Green's functions while the function $B_{L,N}$ can be expanded in anti-time ordered Green's functions. The terms in the expansion are called half-diagrams and the
sub-index $\irr$ in \Eq{eq:A_def} and \Eq{eq:B_def} indicates that 
we remove all half-diagrams that will lead to a reducible self-energy by a gluing procedure that we will describe in more detail later; this also implies that the sum over particle-hole pairs in (\ref{eq:sigless2}) starts at $N=1$. The diagrammatic expansion of $A_{L,N}$ and $B_{L,N}$ is most easily performed using the known Feynman rules
(see \ref{app:prefactors}) of the contour-ordered $n$-particle Green's function, which is defined as \cite{Stefanucci2013}
\begin{equation} \label{eq:GN}
    G_n (1, \ldots n; 1' \ldots n') =
    \frac{1}{\rmi^n} \langle \mathcal{T}_\gamma [ \hat{c}_{1,H} (z_1) \ldots 
    \hat{c}_{n,H} (z_n)\hat{c}^\dagger_{n',H} (z_n') \ldots \hat{c}^\dagger_{1',H} (z_1') ] \rangle
\end{equation}
where $\hat{c}_{j,H}$ and $\hat{c}_{j,H}^\dagger$ are the Heisenberg forms of $\hat{c}_j$  and $\hat{c}_j^\dagger$. Using this definition
we obtain the expressions
\begin{align}
    A_{L,N} (1) &= 
    \rmi^{N+2} \sum_{j_1 j_2 j_3} \gamma_{j_1 j_2 j_3} (\bfx_1) 
    G_{N+2}^T (j_2,j_3,i_1, \ldots i_N; j_1^+, i_1^\prime, \ldots , i_{N+1}^\prime) \label{eq:A_Gn}\\
    B_{L,N} (2) &= 
    \rmi^{N+2} \sum_{j_1 j_2 j_3} \gamma_{j_1 j_2 j_3}^* (\bfx_2) 
    G_{N+2}^{\bar{T}}( j_1^+, i_1^\prime, \ldots , i_{N+1}^\prime ;
    j_2,j_3,i_1, \ldots i_N) \label{eq:B_Gn}
\end{align}
where all the operators with indices in $I$ and $I'$ have time-coordinate $T$ and the operators with labels in the set $\{ j_1, j_2, j_3)$ have time $t_1$ in $A_{L,N}$ and $t_2$ in $B_{L,N}$. The Green's function $G_{N+2}^T$ denotes an $(N+2)$-particle Green's function ordered on the forward contour $\gamma_-$ while $G_{N+2}^{\bar{T}}$ denotes an $(N+2)$-particle Green's function ordered on the backward contour $\gamma_+$.
This gives the diagrammatic expression
\begin{equation} \label{eq:sigless3}
    - \rmi \Sigma_c^< (1,2) =
    \sum_{N=1}^\infty  \frac{(-1)^{N+1}}{N! (N+1)!} \sum_L
    \left[  \bigdiagramcenter{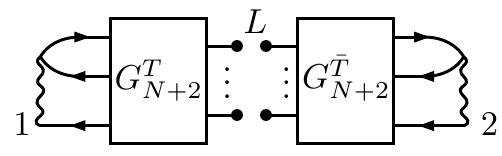} \right]_\irr 
\end{equation}
where the $(-1)^{N+1}$ arises from multiplication of the two factors $\rmi^{N+2}$ in \Eq{eq:A_Gn} and \Eq{eq:B_Gn} along with an additional prefactor $-1$ that arises from assigning a factor $\rmi$ for the two explicitly drawn interaction lines in the bracket to be able to use the Feynman rules for the two amputated $(N+1)$-particle Green's functions that emerge after joining of the $\hat{\gamma}$ operators. In \ref{app:prefactors} we specify the precise Feynman rules for the diagrammatic terms in brackets and demonstrate that a gluing procedure leads to the standard Feynman rules for the self-energy \cite{Stefanucci2013}.


The factorization of the self-energy into half-diagrams in equations \Eq{eq:sigless2} and \Eq{eq:sigless3} was the crucial starting point of the PSD perturbation theory for positive spectra in our previous work \cite{Stefanucci2014}. Let us now investigate whether
we can generalize this derivation to the case of finite temperature systems.
Instead of using the Gell-Mann-Low theorem we
follow Keldysh \cite{Keldysh1965} in making the adiabatic assumption
\begin{equation} \label{eq:adiabatic_assumption}
    \rhoh = \hat{U}_\eta(t_0,-T) \rhoh_0 \hat{U}_\eta(-T,t_0),
\end{equation}
where $\rho_0$ is the density operator of a noninteracting system and $\hat{U}_\eta$ includes an adiabatic switch-on of the interactions from a distant time $-T$ in the past with $\eta$ an adiabatic parameter. 
This is a much stronger assumption than the Gell-Mann-Low theorem as it assumes that all eigenstates are adiabatically connected and that no level crossings occur that lead to degeneracies \cite{Stefanucci2013}. For the moment we explore the consequences of this assumption but we will demonstrate later that the same results can be derived under weaker assumptions. 
Under the adiabatic assumption the lesser self-energy takes the form
\begin{equation}
-\rmi\Sigma_c^<(1,2) = \trace{\rhoh_0 \Uh_\eta(-T, t_0) \hat{\gamma}^\dagger_H (2)  \hat{\gamma}_H (1) \Uh_\eta(t_0, -T)}_{\irr}.
\end{equation}
We introduce a suitable basis of non-interacting many-body eigenstates $| J \rangle$ to perform the trace and place a completeness relation for a complete set of states $| L \rangle$ between the $\hat{\gamma}$-operators. This allows the density-matrix, expressed as $\rhoh_0 = \frac{e^{-\beta(\Hh_0 - \mu \Nh)}}{Z_0}$, to be brought outside the trace, leading to an expression for the self-energy of the form
\begin{equation} \label{eq:S_finite_temperature_2}
-\rmi\Sigma^<_c(1,2) = 
\sum_{L,J} \frac{e^{-\beta (E_J - \mu N_J)}}{Z_0} \la J |  \cop{\gamma}{H}(2) | L \rangle_{\irr} \langle L | \aop{\gamma}{H}(1) | J \ra_{\irr}.
\end{equation}
We could now try and follow the derivation for the zero-temperature case to expand the Lehmann amplitudes $\langle L | \aop{\gamma}{H}(1) | J \rangle$ in Feynman diagrams. However, these can not be expressed straightforwardly in terms of finite temperature Green's functions as the amplitudes neither involve traces over a density matrix nor satisfy appropriate Kubo-Martin-Schwinger boundary conditions. Instead the zero-temperature Wick theorem could be used to expand in diagrams for a zero-temperature many-particle Green's function, but this procedure turns out to be cumbersome and we will therefore follow a much more direct alternative approach.

We first expand $\Sigma$ in Feynman diagrams with finite-temperature Green's functions for all its $g$-lines. Then to connect to the Lehmann expression above we expand each diagram in terms of the zero-temperature Green's functions by writing $g=g_0+ \delta g$ where $g$ is the non-interacting Green's function at finite temperature and $g_0$ is the non-interacting Green's function at zero temperature. Their difference $\delta g$ can conveniently be expressed in terms of $g_0$ along with Fermi factors. We illustrate the procedure with an example.

Let us consider the lesser component of the following diagram that appears in the $T$-matrix approximation for the self-energy
\begin{equation}
D^<(1,2) = \left[ \bigdiagramcenter{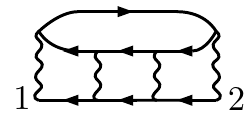} \right]^<
\end{equation}
in which the Green's function lines now denote finite temperature Green's functions. This component of the self-energy is obtained by taking $z_1$ to be on the forward branch $\gamma_-$ and $z_2$ to be on the backward branch $\gamma_+$.
Its expansion in terms of (anti)time-ordered Green's functions is obtained by splitting each internal contour-time integral explicitly into its forward and backward parts, giving
\begin{equation} \label{eq:lehmann_example}
    D^<(1,2) = \bigdiagramcenter{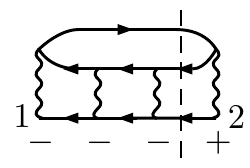} + \bigdiagramcenter{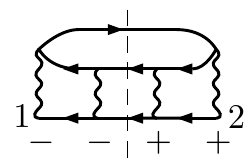} + \bigdiagramcenter{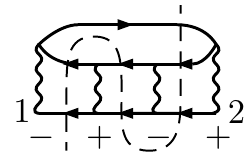} + \bigdiagramcenter{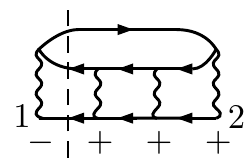}
\end{equation}
where $-$/$+$ denotes that the time-argument is taken to be on the forward/backward branch, and the dashed line marks the separation between the two branches. The Green's functions connecting times on $\gamma_-$ and $\gamma_+$ are either time-ordered ($g_{--}$) and
anti-time-ordered ($g_{++}$) or given by  $g_{+-}=g^>$ and $g_{-+}=g^<$. 
These are the only Green's functions that appear since the two-body interaction is only present in the evolution operators on the real time axis and no interactions occur on the Matsubara branch as we replaced the density matrix by $\hat{\rho}_0$ using the adiabatic assumption \cite{Keldysh1965}.

To connect the self-energy to its Lehmann expansion we write the contour-ordered single-particle Green's function on $\gamma_-$ and $\gamma_+$ at finite temperature as
\begin{equation} \label{eq:g_T_correction}
    g_i(z_1,z_2) = g_{0,i}(z_1,z_2) + \delta g_i(t_1,t_2),
\end{equation}
where $i$ labels a state in the eigenbasis of the one-body Hamiltonian and $g_{0,i}$ is the zero-temperature Green's function 
\begin{equation} \begin{split} \label{eq:contour_ordered_g}
  g_{0,i}(z_1,z_2) &= \theta(z_1,z_2) g^>_{0,i}(t_1,t_2) + \theta(z_2,z_1) g^<_{0,i}(t_1,t_2) \\
  &= \theta(z_1,z_2) (-\rmi) \bar{n}_i \rme^{-\rmi \epsilon_i (t_1 - t_2)} + \theta(z_2,z_1) \rmi n_i \rme^{-\rmi \epsilon_i (t_1 - t_2)},
\end{split} \end{equation}
where $n_i$ is the occupation number of one-particle state $i$ and $\bar{n}_i = 1 - n_i$.
The finite temperature correction term $\delta g_i$ can be expressed as
\begin{equation} \begin{split} \label{eq:g_correction_term}
\delta g_i(t_1,t_2) &= \rmi(f_i - n_i) \rme^{-\rmi \epsilon_i (t_1 - t_2)} = \rmi \left[ \bar{n}_i f_i - n_i \bar{f}_i \right] \rme^{-\rmi \epsilon_i (t_1 - t_2)} \\
&= - f_i\, g^>_{0,i}(t_1,t_2) - \bar{f}_i\, g^<_{0,i}(t_1,t_2),
\end{split} \end{equation}
where $f(\omega) = \frac{1}{\rme^{\beta \omega} + 1}$ and $\bar{f}(\omega) = 1 - f(\omega)$ are the Fermi- and anti-Fermi-functions, and we have used the shorthand notation $f_i = f(\epsilon_i - \mu)$ and similarly for $\bar{f}_i$.
The correction terms do not involve any temporal step-functions, and consequently the same correction applies to the greater and lesser components individually.
This allows us to express the finite temperature diagrams as linear combinations of zero-temperature diagrams weighted by Fermi factors. 
In particular using \Eq{eq:g_T_correction} and \Eq{eq:g_correction_term}, along with the relations 
\begin{equation} \begin{split} \label{eq:g_T_cutting}
    \rmi g^<(t_1,t_2) &= g_{--}(t_1,T) g_{++}(T,t_2) = g_{++}(t_1,-T) g_{--}(-T,t_2) \\
    -\rmi g^>(t_1,t_2) &= g_{++}(t_1,T) g_{--}(T,t_2) = g_{--}(t_1,-T) g_{++}(-T,t_2),
\end{split} \end{equation}
we can write for example the greater component of the finite temperature Green's function as
\begin{equation} \begin{split} \label{eq:g_time_ordered_split}
    -\rmi g^>_i(t_1,t_2) &= g_{0,i,++}(t_1,T) g_{0,i,--}(T,t_2) \\
    &+ f_i\, g_{0,i,++}(t_1,T) g_{0,i,--}(T,-T) g_{0,i,++}(-T,T) g_{0,i,--}(T,t_2) \\
    &- \bar{f}_i g_{0,i,++}(t_1,-T) g_{0,i,--}(-T,T) g_{0,i,++}(T,-T) g_{0,i,--}(-T,t_2),
\end{split} \end{equation}
which can be expressed diagrammatically as (using a double-line to denote the finite temperature $g$-line)
\begin{equation} \label{eq:g_correction_diag}
    -\rmi \bigdiagramcenter{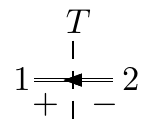} = \bigdiagramcenter{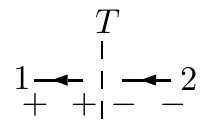} + f_i \bigdiagramcenter{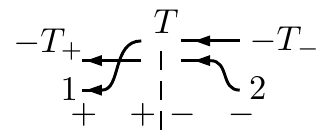} + \bar{f}_i \bigdiagramcenter{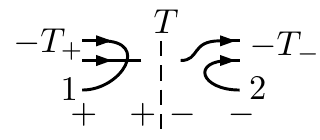}
\end{equation}
where the cut is made at time $T$.
Note that the last two diagrams here belong to a two-particle Green's function, and the differences in signs of the terms appearing in \Eq{eq:g_time_ordered_split} are absorbed into the corresponding diagram prefactors. Diagrammatically the correction term \Eq{eq:g_correction_term} therefore generates two additional diagrams for each $g$-line. These additional diagrams can be interpreted to result from the fact that one can not differentiate between excitations created by interaction with the propagating particle, and those present in the finite temperature ensemble state. Therefore we need to include the processes in which the propagating particle/hole is either exchanged with an ensemble particle/hole excitation or combines with an ensemble hole/particle.

If we consider, for example, the second diagram in \Eq{eq:lehmann_example} and insert \Eq{eq:g_correction_diag} into one of the $g$-lines which connect the $+$ and $-$ halves of the diagram, we generate the diagrams
\begin{equation} \label{eq:lehmann_cut_example}
    -\rmi \bigdiagramcenter{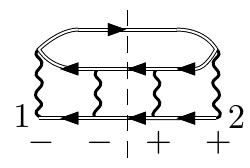} = \bigdiagramcenter{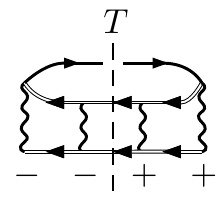} + f_i \bigdiagramcenter{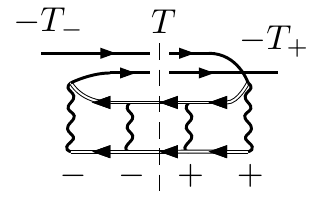} + \bar{f}_i \bigdiagramcenter{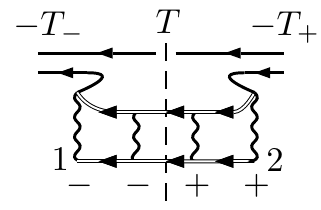}.
\end{equation}
By repeating these steps for each of the $g$-lines crossing the cut in \Eq{eq:lehmann_cut_example} one can achieve a factorization in terms of fully time-ordered and anti-time-ordered half-diagrams at finite temperature.

In an analogous manner the internal finite temperature $g$-lines in the half-diagrams can also be expanded in zero-temperature $g$-lines. Suppose, for example, that we have factorized all the connecting $g$-lines in \Eq{eq:lehmann_cut_example} and we take the diagram with no coupling to the ensemble excitations. An internal $g$-line can be expanded as
\begin{equation} \label{eq:lehmann_cut_example_2}
    \bigdiagramcenter{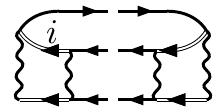} = \bigdiagramcenter{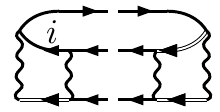} + f_i \bigdiagramcenter{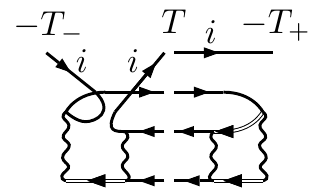} + \bar{f}_i \bigdiagramcenter{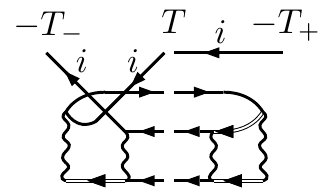}.
\end{equation}

In order to group the half-diagrams into Hermitian products, the Fermi factors need to be expressed in a way that allows them to be taken out as common factors. This can be achieved by noting that for example the diagram multiplied by $f_i$ in \Eq{eq:lehmann_cut_example_2} contains $g^>_i$ and is therefore non-zero only when the state $i$ is unoccupied in the ground-state, i.e. $\bar{n}_i = 1$. We can thus write the prefactor as $f_i = \bar{n}_i f_i + n_i \bar{f}_i$. Similar reasoning works also for the diagram with prefactor $\bar{f}_i$ in \Eq{eq:lehmann_cut_example_2}, which allows us to pull $\bar{n}_i f_i + n_i \bar{f}_i$ out as a common factor. 

We could then, in theory, construct PSD self-energy approximations by expanding the $g$-lines connecting the half-diagrams, and building Hermitian products such as
\begin{equation} \begin{split} \label{eq:psd_example}
    -\rmi \Sigma^<(t_1,t_2) &= \bigdiagramcenter{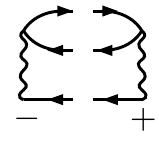} + \frac{\bar{n}_i f_i + n_i \bar{f}_i}{2} \left[ \bigdiagramcenter{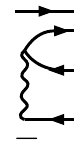} + \bigdiagramcenter{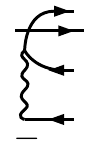} \right] \left[ \bigdiagramcenter{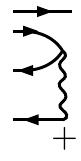} + \bigdiagramcenter{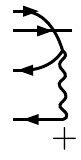} \right] + \ldots \\
    &= -\rmi\bigdiagramcenter{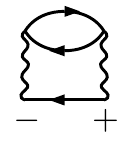} + (\bar{n}_i f_i + n_i \bar{f}_i) \left[ \bigdiagramcenter{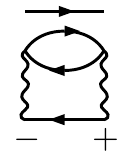} + \bigdiagramcenter{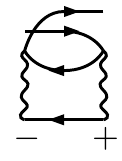} \right] + \ldots \,.
\end{split} \end{equation}

Terms such as in \Eq{eq:psd_example} can be connected to the Lehmann amplitudes appearing in \Eq{eq:S_finite_temperature_2} by noting that for example a $g^>_i$ line starting at $-T$ corresponds to a state $| J \ra$ in \Eq{eq:S_finite_temperature_2} with the single-particle state $i$ occupied. An excitation to one-particle state $i$ can only occur in $| J \rangle$ when that state is unoccupied in the ground state. Similarly
a de-excitation from one-particle state $i$ can only occur in $| J \rangle$ 
when that state is occupied in the ground state. 
We can therefore express the prefactor as $\bar{n}_i f_i + n_i \bar{f}_i = n^J_i f_i + \bar{n}^J_i \bar{f}_i$, where $n^J_i$ is the occupation of the single-particle state $i$ in the non-interacting many-body eigenstate $|J\ra$. In general such terms appear as products, which can be expressed as
\begin{equation}
    \prod_i \left[ n^J_i f_i + \bar{n}^J_i \bar{f}_i \right] 
    = \prod_i \frac{\rme^{-\beta n^J_i \epsilon_i}}{1 + \rme^{-\beta \epsilon_i}} 
    = \frac{\rme^{-\beta \sum_i n^J_i \epsilon_i}}{ \prod_i (1 + \rme^{-\beta \epsilon_i})},
\end{equation}
and thus be related to the Boltzmann factor for the state $|J\ra$ appearing in \Eq{eq:S_finite_temperature_2}.
In this manner one could, by careful accounting of all the diagrams and prefactors, work one's way back to a diagrammatic expression of the form of \Eq{eq:S_finite_temperature_2}, which would constitute a finite temperature generalization of \Eq{eq:sigless3} containing additional connections between the half-diagrams that are weighted by Boltzmann factors corresponding to the sum over the $J$-states. This is equivalent to defining the $J$ and $L$ states in \Eq{eq:S_finite_temperature_2} in a suitable manner to write the Lehmann amplitudes $\langle J | \cdots | L \rangle_{\irr}$ as multi-particle Green's functions at zero temperature.
We have therefore succeeded in deriving a diagrammatic expansion for the self-energy based on Lehmann amplitudes which generalizes our earlier work and reduces to it in the
zero temperature limit. The finite temperature corrections of this expansion can be interpreted as additional interactions by which particles enter and leave a heat bath \cite{Landshoff1996}.  
Although physically insightful and feasible to derive a PSD perturbation theory, the approach
is not very practical; as follows from \Eq{eq:g_correction_diag} each cut $g$-line leads to three zero-temperature diagrams. In the following section we discuss a much more viable approach to PSD approximations at finite temperature.

\subsection{Factorization of the self-energy and the issue of non-vanishing vacuum diagrams}
\label{sec:factorization}

In the previous section we provided a generalisation of the zero-temperature cutting procedure that
can be directly related to the Lehmann amplitudes $\langle L | \gamma (1) | J \rangle_{\irr}$.
However, this required a formulation in terms of zero temperature Green's functions and 
is, although possible \cite{Blazek2021,Blazek2022}, not very practical for deriving PSD approximations for applications of many-body perturbation theory. We therefore advance here another procedure for factorizing self-energy diagrams.

We consider again the standard finite temperature diagrammatic expansion of the self-energy from the previous section. We expand each diagram in (anti)time ordered components as in \Eq{eq:lehmann_example} and then
algebraically factorize the $g$-lines connecting the forward and backward branches using the relation 
\begin{equation}
    g^\lessgtr (t,t') = g^R (t,t_0) g^\lessgtr (t_0, t_0) g^A (t_0,t')
\end{equation}
where $g^R$ and $g^A$ are the retarded and advanced Green's functions and $t_0$ is a suitably chosen time. Since $g^<(t_0,t_0)=\rmi \rho_0$ and $g^> (t_0,t_0)=-\rmi (1- \rho_0)=-\rmi \bar{\rho}_0$ where $\rho_0$
is the one-particle density matrix at time $t_0$, we can write
\begin{align}
    -\rmi g^<(t,t') &= [ g^R (t,t_0) \rho_0^{\frac{1}{2}} ] [ g^R (t',t_0) \rho_0^{\frac{1}{2}} ]^\dagger = \tilde{g}^< (t,t_0)  [ \tilde{g}^< (t',t_0) ]^\dagger 
    \label{eq:tilde_less}\\
    \rmi g^> (t,t') &= [ g^R (t,t_0) \bar{\rho}_0^{\frac{1}{2}} ] [ g^R (t',t_0) \bar{\rho}_0^{\frac{1}{2}} ]^\dagger  =
    \tilde{g}^> (t,t_0)  [ \tilde{g}^> (t',t_0) ]^\dagger
    \label{eq:tilde_great}
\end{align}
where the $\tilde{g}^\lessgtr$ are defined by these equations and
$\rho_0^{\frac{1}{2}}$ and $\bar{\rho}_0^{\frac{1}{2}}$ are the square roots of $\rho_0$ and
$\bar{\rho}_0$ as spatial operators, which are well-defined since these operators are PSD;
in case we use the eigenbasis of the one-body Hamiltonian of (\ref{eq:hamiltonian}) they are diagonal matrices with diagonal elements given by the square roots of the (anti)-Fermi functions. We thus see that we can factorize the lesser and greater functions as
$\mp \rmi g^\lessgtr = \tilde{g}^\lessgtr \tilde{g}^{\lessgtr \dagger}$ which, as shown in \ref{app:prefactors}, allows the self-energy to be written as
\begin{equation}
    -\rmi \Sigma^<_c (1,2) = \sum_{N=1}^\infty   \sum_{a,b,I,P} (-1)^{|P|} A_{N,I}^{(a)}(1,t_0)  A_{N,P(I)}^{(b)*} (2,t_0)
    \label{eq:AAP}
\end{equation}
where $A^{(a)}_{N,I}$ represents a diagrammatic expression of a half-diagram of topology $a$. 
The label $N$ denotes the number of particle-hole line pairs connecting the half-diagrams and $I$ is a multi-index for the spatial labels of all connecting lines. 
The permutations $P$ run over all permutations of the particle and hole lines separately and since for $\Sigma^<$ there is one hole line more than a particle line there are $N!(N+1)!$ of such permutations.
The main difference with the derivation in the previous section is that the diagram $A^{(i)}_{N,I}(1,t_0)$
can not directly be associated with a Lehmann amplitude; it rather consists of a term in an expansion of the $(N+1)$-particle Green's function in which the Green's functions on cut external legs have been replaced by the functions $\tilde{g}^\lessgtr$, and the remaining leg corresponding to vertex $1$ has been removed. In \ref{app:prefactors} we prove that we recover the correct expansion of the self-energy when these modified diagrams obey the same Feynman rules as the $(N+1)$-particle Green's function. From expression (\ref{eq:AAP}) it also becomes clear that the expansion is PSD; defining
\begin{equation}
    S_{N,I}^{(a)} (1,t_0) = \sum_P (-1)^{|P|} \, A^{(a)}_{N,P(I)} (1,t_0)
\end{equation}
we see that the self-energy can be written as the explicitly positive semi-definite expression
\begin{equation}
    -\rmi \Sigma^<_c (1,2) = \sum_{N=1}^\infty \frac{1}{N!(N+1)!} \sum_{a,b,I} S_{N,I}^{(a)} (1,t_0) S_{N,I}^{(b)*} (2,t_0).
    \label{eq:ssn}
\end{equation}

So far the discussion concerned the exact self-energy. However, in contrast to the exact self-energy,
a given diagrammatic approximation, such as that in \Eq{eq:lehmann_example}, will in general not lead to a positive definite spectral function. To solve this problem we can attempt to repeat the procedure for the zero temperature
case \cite{Stefanucci2014} to construct a PSD perturbation series.
The first step is to write the expansion as a product of factors containing the time and anti-time-ordered parts;
again using our factorization procedure with the functions $\tilde{g}$ of equations \Eq{eq:tilde_less} and \Eq{eq:tilde_great}.
%
%
The factors that result from this procedure have then to be re-assembled into Hermitian products
to obtain the PSD form of \Eq{eq:ssn}.
For systems at finite temperature a complication arises from
the presence of diagrams with islands consisting solely of time or anti-time ordered vertices of the form
\begin{equation} \label{eq:islands}
 \bigdiagramcenter{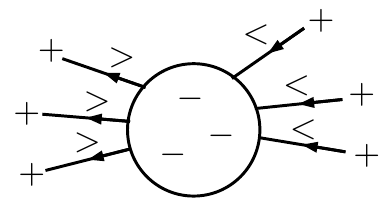}
\end{equation}
such as in the third diagram in \Eq{eq:lehmann_example}. At zero-temperature such islands vanish, as demonstrated by the following argument. All the $g$-lines entering an island will be greater/lesser, and all the $g$-lines leaving will be lesser/greater. At zero temperature a greater/lesser Green's function always carries an energy above/below the Fermi level and the island formally has a net energy flow in or out. This is, however, forbidden by energy conservation that is mathematically enforced by the presence of Heaviside functions whereby the island vanishes. 
At finite temperature this is not true anymore due to the smearing of the Heaviside functions to continuous Fermi functions. This is problematic when it comes to constructing PSD approximations out of half-diagrams,
which is best illustrated with the example of the third diagram in \Eq{eq:lehmann_example}.
The diagram has the following structure
\begin{equation}
    \bigdiagramcenter{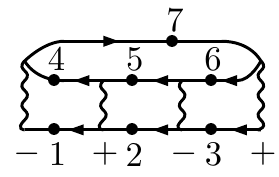} = \bigdiagramcenter{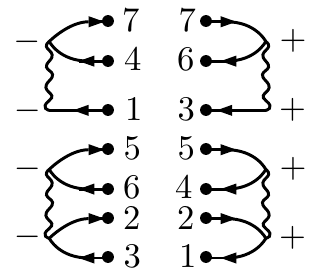},
\end{equation}
which we have cut into time and anti-time ordered parts using the factorization of the greater and lesser Green's functions, and where two islands appear that are disconnected from external vertices. In the notation of \Eq{eq:AAP} the diagrammatic product on the right-hand side corresponds to the term
\begin{equation}
    (-1)^2 A^{(a)}_{3,(3265147)} (1,t_0) A^{(a)*}_{3,(1245367)} (2,t_0)
\end{equation}
where the half-diagrams on both sides of the cut are of the same topology $a$, have cut lines with $N=3$ particle hole pairs and
cut labels $I=(3265147)$ and $P(I)=(1245367)$ corresponding to an even permutation $(13)(46)$ consisting of two transpositions.
In order to obtain a PSD self-energy containing this diagram we need to construct a PSD form from the separate half-diagrams as derived in our earlier work \cite{Stefanucci2014}. If the half-diagrams have the same topology and only differ in the labeling of the cut legs, this is achieved by adding the half-diagrams corresponding to the smallest permutation subgroup containing the permutation of the legs in question. 
This procedure amounts to the construction of additional terms in \Eq{eq:AAP} which allows for a rewriting of the equation in a PSD form as is done in \Eq{eq:ssn}.
In our case the halves differ by
the permutation $\sigma=(13)(46)$ of the external legs and the smallest subgroup containing it is $\{ \iota, \sigma \}$ where $\iota$ is the identity permutation. This yields the following minimal PSD extension of the given diagram
\begin{equation}
 \bigdiagramcenter{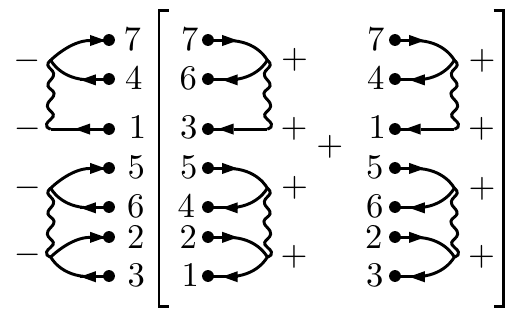} = \bigdiagramcenter{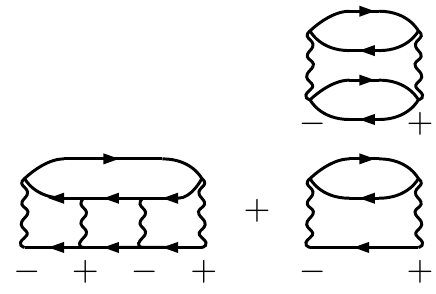}
 \label{bothersome_bubble}
\end{equation}
We therefore see that the construction introduces a new self-energy diagram containing a disconnected vacuum diagram.
This is readily seen to be a general feature; 
the minimal PSD extension of diagrams containing $+$ or $-$ islands necessarily introduces disconnected vacuum diagrams into the self-energy. However, such vacuum diagrams do not contribute to the exact self-energy, since a sum over all their different time/anti-time ordered components corresponds to an integral over the loop contour on all vertices, which vanishes as a closed integral. The minimal PSD extension procedure, however, creates only specific time-ordered components of the vacuum diagram, and therefore introduces a finite contribution that would eventually be cancelled in the exact expansion. 
It is clearly undesirable to include in an approximate expansion terms that should ultimately not contribute. This issue has been prominently discussed in the particle physics literature \cite{Kobes1985,Jeon1993,Landshoff1996,Bedaque1997,Gelis1997} where diagrams containing $+$ or $-$ islands are referred to as "non-cuttable" diagrams \cite{Kobes1986,Gelis1997}.
In section \ref{eq:retarded_cutting_rules} we will resolve this issue using the concept of retarded half-diagrams.

However, before addressing that problem we will first derive a generalization of the result of this section
that is even valid for non-equilibrium final states and does not require the adiabatic assumption. 
It will also allow us to make a connection to earlier work of Jeon \cite{Jeon1993} which is thereby put into a different context.

\subsection{Time-ordered cutting rules from the Riemann-Lebesque lemma}
\label{sec:time_ordered_cutting}

The derivation of the finite temperature expansion for the self-energy in section \ref{sec:cutting_lehmann} was based on the assumption that the interacting density matrix can be adiabatically connected to a non-interacting one \cite{Keldysh1965}. This assumption is stronger than needed and in this section we derive the
result in a more direct and simpler fashion by applying the Riemann-Lebesque (RL) lemma for Fourier transforms to the case of systems with a continuous one-particle spectrum.
In the second part of this section we will show that even the assumption
of a continuous spectrum is not needed, at least when we confine ourselves to finite temperature systems in thermodynamic equilibrium.

Our starting point is the full contour of Figure \ref{fig:contour} which contains also a Matsubara branch.
Throughout this section we use as an example the diagram
\begin{equation} \label{eq:T_matrix_example_1}
D^<(1,2) = \left[ \bigdiagramcenter{T_matrix_limit} \right]^<
\end{equation}
The lesser component is obtained by taking the external time $z_1$ on the forward branch $\gamma_-$ and the external time $z_2$
on the backward branch $\gamma_+$.
The time-ordered expansion of a contour diagram at finite temperature is achieved by splitting each internal contour-time integral explicitly into its forward, backward and Matsubara parts (see Figure \ref{fig:contour}) 
\begin{equation}
\int_\gamma dz = \int_{\gamma_-} dz + \int_{\gamma_+} dz + \int_{\gamma_M} dz = \int_{t_{0-}}^\infty dz + \int_\infty^{t_{0+}} dz + \int_{t_{0+}}^{t_0 - \rmi\beta} dz
\end{equation}
and summing over all the resulting real-time integrals.
For our example diagram (using $-/+$ to denote the $\gamma_-$/$\gamma_+$ branches)
we have
\begin{equation} \begin{split} \label{eq:time_ordered_expansion}
D^<(1,2) &= \bigdiagramcenter{T_matrix_plusminus_1} + \bigdiagramcenter{T_matrix_plusminus_2} + \bigdiagramcenter{T_matrix_plusminus_3} + \bigdiagramcenter{T_matrix_plusminus_4} \\
&+ \bigdiagramcenter{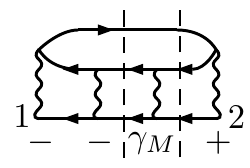} + \bigdiagramcenter{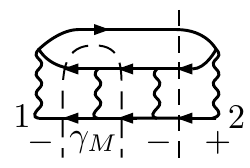} + \bigdiagramcenter{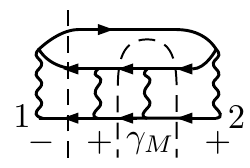} + \bigdiagramcenter{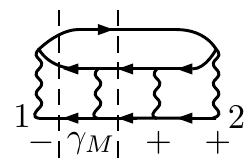} \\
&+ \bigdiagramcenter{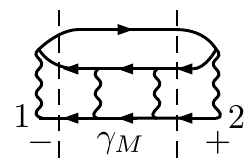}
\end{split} \end{equation}
where dashed lines mark the cuts between the distinct branches.
The Green's functions that connect a real time to a time on the Matsubara branch are
denoted by $g(t_0-\rmi\tau,t_\pm)=g^\lceil (\tau,t)$ and $g (t_\pm,t_0-\rmi\tau)=g^\rceil (t,\tau)$
(see \cite{Stefanucci2013} for a thorough discussion) and a Green's function connecting two times on
the Matsubara branch is denoted by the Matsubara Green's function $g^M (\tau,\tau')$. We will now demonstrate that 
for a system with a continuous energy spectrum the so-called left and right
Green's functions $g^\lceil$ and $g^\rceil$ give a vanishing contribution to the Feynman diagrams provided we assume the system to be in equilibrium before a given time $t_0$. Practically this implies that we can disregard the Matsubara branch in the diagrams provided we let all real time integrals commence at $-\infty$ instead of the finite time $t_0$.

\begin{figure*}
\centering
\includegraphics[scale=1.4]{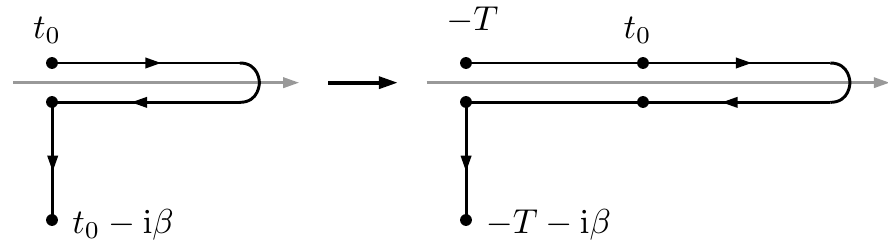}
\caption{In a system initially in equilibrium, the wait time between state preparation and measurement does not affect the result. One can therefore extend the contour to a past time $t = -T$. \label{fig:extension}} 
\end{figure*}

In the following we assume that the system is in equilibrium for times $t \leq t_0$ while for $t>t_0$ there may be external perturbations that bring the system into a non-equilibrium state.
From time translation invariance it follows that for a system in equilibrium the value of an observable cannot depend on the wait time between state preparation and the start of the measurement. Therefore we can extend the contour to a past time $t = -T \leq t_0$ (see Figure \ref{fig:extension}). 
Now for any time $t$ we have the decomposition (see \cite{Stefanucci2013} for details),
\begin{equation} \label{eq:Riemann_Lebesque_argument}
g^\lceil(\tau,t) = -\rmi g^M(\tau,0) g^A(-T,t) \quad \quad g^\rceil (t,\tau) = \rmi g^R (t,-T) g^M (0,\tau)
\end{equation}
where $g^{R/A}$ denotes the retarded/advanced Green's function. For $t > t_0$ it will be useful to use a 
semi-group property \cite{Stefanucci2013} of the retarded and advanced functions
\begin{equation}
    g^A (-T,t) = -\rmi g^A (-T,t_0) g^A (t_0, t)  \quad \quad g^R (t,-T) = \rmi g^R (t,t_0) g^R (t_0,-T).
    \label{eq:semigroup}
\end{equation}
The functions $g^{R/A}$ are in general expressed in terms of time-ordered exponentials \cite{Stefanucci2013}
but for times $t$ and $t'$ in the equilibrium time interval $[-T,t_0]$ they acquire a simple form with respect to a basis of 
non-interacting one-particle eigenstates of the one-body part of the Hamiltonian
\begin{equation}
    g_i^R (t,t') = -\rmi \theta (t-t')  \rme^{-\rmi \epsilon_i (t-t')}  \quad \quad
    g_i^A (t,t') = \rmi \theta (t'-t)  \rme^{-\rmi \epsilon_i (t-t')} 
\end{equation}
where $\epsilon_i$ is the eigenenergy of the one-particle state labeled by $i$.
The key ingredient of our derivation is the Riemann-Lebesque (RL) lemma (see Th.7.5 of \cite{Rudin}) which states that
\begin{equation} \label{eq:RL_lemma}
    0 = \lim_{T \rightarrow \pm \infty} \int \rmd\epsilon \, F(\epsilon) \, \rme^{\rmi \epsilon T} 
\end{equation}
whenever $F$ belongs to the space of integrable functions  $L^1 (\mathbb{R})$ 
(for example $F$ can be a piecewise continuous function). If the one-particle spectrum 
$\epsilon_i$ is continuous, as is generally the case for an infinite system, the summations
over the labels $i$ in the Feynman diagrams can be replaced by integrals over the energy $\epsilon$
where a density of states $d(\epsilon)$ appears in the integration volume element. If $d(\epsilon)$
belongs to $L^1 (\mathbb{R})$ the RL lemma and \Eq{eq:semigroup} can be invoked to establish that
diagrams that contain $g^\rceil$ and $g^\lceil$ vanish in the limit $T \rightarrow \infty$.

Applying the RL lemma we therefore retain only the first row of \Eq{eq:time_ordered_expansion} provided that in these diagrams we let the lower bound of all time integrations be equal to $-\infty$. We see that in this way we derived precisely the same self-energy expansion as in the previous section with two important differences. First, we needed a much weaker assumption (a continuous spectrum instead of the adiabatic assumption). Second, the expression derived above is valid for general systems in initial equilibrium up to some finite time but exposed to a time-dependent external perturbation after it.

In the following we will show that even the assumption of having a continuous one-particle spectrum can be disposed off when we restrict ourselves to systems in thermodynamic equilibrium. This will be done by making a connection to the work of Jeon\cite{Jeon1993}.
We can write the self-energy in an alternative version of \Eq{eq:AAP} as follows
\begin{equation} \label{eq:sigma_time_ordered}
    -\rmi\Sigma_c^<(t_1,t_2) = \sum_{N=1}^\infty \sum_{a,b,I,P} (-1)^{|P|} \int_{\gamma_-} \rmd t_\Ical \int_{\gamma_+} \rmd t'_\Ical\, B_{N,I}^{(a)} (t_1,t_\Ical) \Upsilon_I^<(t_\Ical,t'_\Ical) B_{N,P(I)}^{(b)*} (t_2,t'_\Ical),
\end{equation}
where the terms $B_{N,I}^{(a)}$ are obtained from the terms $A_{N,I}^{(a)}$ by removing the legs of the cut lines and $t_{\Ical}$ and $t_\Ical^\prime$ are sets of time labels for the endpoints of the cut lines. 
The cut lines are assembled in a new term $\Upsilon_I$ of the form 
\begin{equation} \label{eq:big_G}
    \Upsilon_{I}^<(t_\Ical,t'_\Ical) = \prod_{p\in J}^N 
    -\rmi g_{p}^>(t'_p, t_p) \prod_{q \in K}^{N+1}  \rmi g_{q}^<(t_{q}, t_{q}^\prime) 
\end{equation}
where the index set $I=(J,K)$ for the cut lines has been divided into sets of indices $J$ referring to particle lines and indices $K$ referring to hole lines (we always choose a one-particle basis in which the zeroth-order Green's functions are diagonal).  

If we specialize to systems that are in equilibrium at all times then all quantities are time-translation invariant and we can do Fourier transforms to find lower dimensional expressions in frequency space.
Fourier transforming \Eq{eq:sigma_time_ordered} one obtains
\begin{equation} \label{eq:sigma_time_ordered_freq}
    -\rmi\Sigma_c^<(\omega) = \sum_{N=1}^\infty \sum_{a,b,I,P} (-1)^{|P|} \int \frac{\rmd\omega_\Ical}{(2\pi)^I} B_{N,I}^{(a)} (\omega_\Ical) 
    2\pi \delta (\omega - \Omega)\Upsilon_I^<(\omega_\Ical) B_{N,P(I)}^{(b)*} (\omega_\Ical),
\end{equation}
where we defined the total energy flowing to the left through the cut lines as $\Omega = \sum_{i \in \Ical} \omega_i$ and
\begin{equation} \label{eq:B_TI-t}
    B_{N,I}^{(a)} (\omega_\Ical) = \int \rmd t_\Ical \, B_{N,I}^{(a)}(t_1,t_\Ical) \rme^{\rmi \omega_\Ical \cdot (t_1-t_\Ical)} 
\end{equation}
(using the notation $\omega_\Ical \cdot (t_1 -t_\Ical) = \sum_{i \in \Ical} \omega_i (t_1- t_i)$) so that the time-convolutions between the half-diagrams are replaced by simple products. 
The function $\Upsilon_I^< (\omega_\Ical)$ with the explicit form
\begin{equation}
\Upsilon^<_I (\omega_\Ical) = \prod_{p\in J}^N 
    -\rmi g_{p}^>(-\omega_p) \prod_{q \in K}^{N+1}  \rmi g_{q}^<(\omega_q) 
\end{equation}
is a PSD integral operator in spatial indices for all $\omega_\Ical$.
If we consider the explicit form of $g^\lessgtr$ (see \Eq{eq:contour_ordered_g}) the Fourier transform of $\Upsilon^<$ is readily expressed in terms of delta distributions and with the relations above we can express the lesser self-energy as
\begin{equation} \begin{split} \label{eq:jeon_result}
   -\rmi \Sigma^<_c(\omega) 
   =\sum_{N=1}^\infty \sum_{a,b,I,P} (-1)^{|P|} B_{N,I}^{(a)} (\tilde{\epsilon}_\Ical) 
    2\pi \delta (\omega - \mathcal{E}) F^<(\epsilon_\Ical) B_{N,P(I)}^{(b)*} (\tilde{\epsilon}_\Ical),
\end{split} \end{equation}
where we defined $\epsilon_\Ical$ the set of one-particle energies $\epsilon_i$ for the cut lines and we further defined
\begin{equation}
F^< (\epsilon_\Ical) = \prod_{p \in J}^N \bar{f}(\epsilon_p - \mu) \prod_{q \in K}^{N+1} f (\epsilon_q - \mu) .
\end{equation}
We also defined the energy $\mathcal{E}=\sum_{q \in K} \epsilon_q- \sum_{p \in J}\epsilon_p$ and $\tilde{\epsilon}_\Ical$
is an argument list consisting of the energies $\epsilon_q$ for the hole lines and $-\epsilon_p$
for the particle lines.
Expression (\ref{eq:jeon_result}) is of a physically appealing Fermi golden rule form \cite{Danielewicz1990}, involving products of scattering amplitudes $B_{N,I}^{(i)}$,
occupation functions incorporated in $F^<$ and an overall energy conservation enforced by a delta function.
A rearrangement of the expression as in \Eq{eq:ssn} will make this even more apparent but the expression above is
closer to the expressions that would actually be used in practice.  

In equilibrium the fluctuation dissipation theorem gives that $-\rmi\Sigma^<_c (\omega) = f(\omega - \mu) \Gamma (\omega)$ and 
$\rmi\Sigma^>_c (\omega) =  \bar{f}(\omega - \mu) \Gamma (\omega)$
where $\Gamma (\omega)$ is the spectral function of the self-energy (also known as the rate function) which can be expressed as
\begin{equation}
\Gamma (\omega) = \rmi(\Sigma_c^>(\omega) - \Sigma_c^<(\omega) ) =\rmi ( \Sigma_c^R(\omega) - \Sigma_c^A(\omega)) .
\end{equation}
If we use that in equilibrium (see for example \cite{Stefanucci2013} section 9.5)
\begin{equation} \label{eq:S_M_S_R_relation}
\Sigma_c^{R/A}(\omega) = \Sigma_c^M(\omega - \mu \pm \rmi\eta),
\end{equation}
with $\eta$ a positive infinitesimal and $\mu$ the chemical potential we can write 
\begin{equation}
-\rmi \Sigma_c^<(\omega) = \rmi f(\omega-\mu)  (\Sigma_c^M(\omega - \mu + \rmi\eta) - \Sigma_c^M(\omega - \mu - \rmi\eta)) .
\label{eq:less_M}
\end{equation}
This relation was used by Jeon \cite{Jeon1993} to derive an expression of the form of \Eq{eq:jeon_result} (for the Green's function instead of the self-energy)
using an expansion in Matsubara Green's functions (see equations (3.4a) and (3.4b) in \cite{Jeon1993}). Jeon's derivation proceeds via the following steps:
\begin{enumerate}
    \item Each Matsubara diagram is expanded in its time-ordered components in imaginary time.
    \item Each time-ordered component is expressed in frequency space using the rules derived originally by 
    Balian and De Dominicis \cite{Balian1960} and later by Baym and Sessler \cite{Baym1963}.
    \item For each of the obtained expressions the limit $\eta \rightarrow 0$ is taken for a difference of terms as in \Eq{eq:less_M}.
    \item The final expression obtained this way can be interpreted as a sum over cut diagrams and factors to the left and right of the cut can be expressed in terms of time-ordered and anti-time-ordered Green's functions (this uses \Eq{eq:time_ordered_equation} of \ref{app:spectral_representation}).
\end{enumerate}
Importantly this derivation nowhere uses the Riemann-Lebesque lemma, and therefore does not need to assume continuous spectra. 
Its main ingredient is analytic continuation and the use of fluctuation dissipation relations such as $g^>(\omega)=-\rme^{\beta (\omega-\mu)} g^< (\omega)$. It is therefore similar in spirit to the derivation that proves that the equilibrium limit of the Kadanoff-Baym equations is independent of the initial time of time-propagation (section 9.6 of \cite{Stefanucci2013}). It follows from Jeon's derivation that in equilibrium the removal of Matsubara track integrals in the expansion of $\Sigma_c^\lessgtr$ is justified even for systems with a discrete energy spectrum provided we extend the time-integrals to the infinite past. The information about the initial density matrix is not lost, as it is still preserved in the functions $g^\lessgtr$ that contain the Fermi functions depending on the temperature and the chemical potential. For additional discussion of these relations see \ref{app:spectral_representation}; they play a more important role in the next section in which we use a more elegant approach to derive the result.

The above cutting rules still retain the issues relating to disconnected pieces in half-diagrams leading to unwanted vacuum diagrams, as we discussed below \Eq{eq:islands}. In the following section we propose how this crucial issue can be solved by a different kind of expansion of the contour diagram, replacing time-ordered half-diagrams by retarded ones.

\subsection{Retarded cutting rules}
\label{eq:retarded_cutting_rules}

The problems posed by disconnected pieces in half-diagrams are created because cutting the diagram between the forward and the backward branches limits each half-diagram to a single branch on which vacuum diagrams do not integrate to zero. The most straightforward way to deal with this issue is therefore to start by deforming the contour to a double loop so that it returns to $-T$ between the external time arguments located at the end of each loop as follows
\begin{equation} \label{eq:deformed_contour}
\includegraphics[scale=1.4]{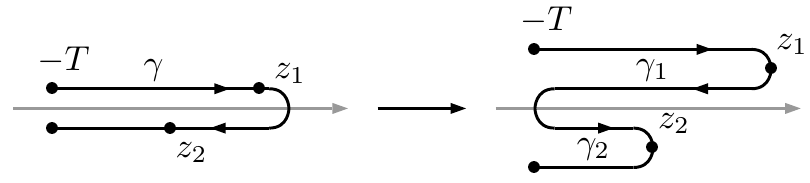}
\end{equation}
Each time integration can therefore be written as a sum of two time integrations on each separate loop.
When this is done for the time-integrals in a Feynman diagram
 disconnected sub-diagrams vanish as they are integrated from $-T$ to $-T$
 on a loop and only the contribution from terms connecting to external vertices survive.
This elimination does not rely on any physical properties of the system, such as energy-conservation, and therefore happens also at finite temperature and in non-equilibrium systems.

Internal loop integrals over a sub-diagram up to the external times representing the entry or exit point of the self-energy can be expressed in terms of real-time integrals over retarded functions \cite{Hyrkas2019,Hyrkas2019b}. Accordingly we define a multi-argument retarded component of a general Feynman diagram with integrand $D(z_\Ncal) = D(z_1, \ldots, z_N)$ and external vertex at time $z_1$ as follows
\begin{equation} \label{eq:retarded_requirement}
\int_{\gamma} \rmd z_2 \cdots \rmd z_N D(z_\Ncal) = \int_{t_0}^\infty \rmd t_2 \cdots \rmd t_N D^{R(1,2 \cdots N)}(t_\Ncal),
\end{equation}
where $\gamma$ is a loop contour and the multi-retarded function $D^{R(1,2 \cdots N)}(t_\Ncal)$ with respect to argument $1$ is a real-time function such that equality holds in the equation above.
The multi-retarded functions allow for convenient conversion from contour to real-time integration and capture, in effect, the back-and-forth integration on the contour within a single real-time function. We show how to evaluate these functions in \ref{app:retarded_diagrams} and \ref{app:spectral_representation}, although 
for the purpose of developing PSD approximations they feature mainly as building blocks and it is sufficient to evaluate the final self-energy directly once it is obtained from the PSD construction.
If we now split our time-integrations as sums of integrations on each of the loops of \Eq{eq:deformed_contour} and on the Matsubara branch $\gamma_M$ and apply this to our example diagram we have
\begin{equation} \begin{split} \label{eq:retarded_expansion}
D^<(1,2) &= \bigdiagramcenter{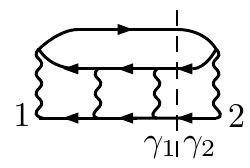} + \bigdiagramcenter{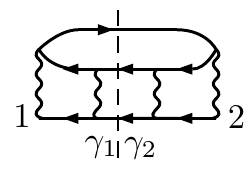} + \bigdiagramcenter{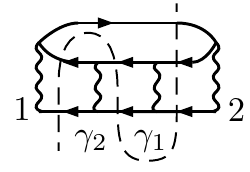} + \bigdiagramcenter{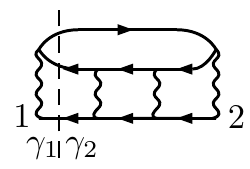} \\
&+ \bigdiagramcenter{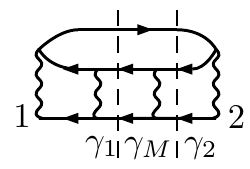} + \bigdiagramcenter{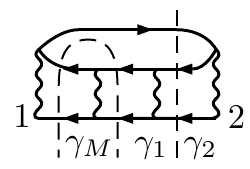} + \bigdiagramcenter{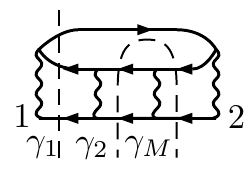} + \bigdiagramcenter{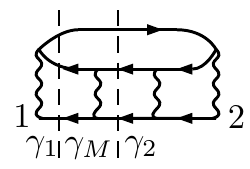} \\
&+ \bigdiagramcenter{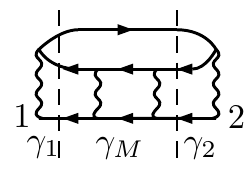},
\end{split} \end{equation}
where in the figure the time integrations over the loops $\gamma_1$, $\gamma_2$ (see Figure \ref{eq:deformed_contour}) and the Matsubara branch $\gamma_M$ are indicated for each interaction line. The lesser component was selected by taking $z_1$ to be on loop $\gamma_1$
and $z_2$ to be on loop $\gamma_2$ which occurs later in contour ordering.
We will assume that our system has a continuous spectrum such that we can apply the Riemann-Lebesque lemma. Then 
as before the Matsubara diagrams can be discarded when time-integrals commence at $-\infty$, so that we retain only the first line of diagrams in \Eq{eq:retarded_expansion}. 
Now since internal vertices are integrated over the separate loops $\gamma_1$ and $\gamma_2$, any half-diagram with disconnected parts vanishes and therefore also the third diagram in the first line of (\ref{eq:retarded_expansion})
is eliminated.
The remaining three diagrams are the same one would obtain in a time-ordered zero temperature treatment, only now with retarded half-diagrams and internal $g$-lines at finite temperature.
The example clearly illustrates the strategy outlined in the beginning of this section, by the use of retarded half-diagrams the bothersome diagram of equation (\ref{bothersome_bubble}) 
no longer appears in the expansion.

Similar to the time-ordered case the self-energy generally takes the form
\begin{equation} \label{eq:sigma_deformation_expansion}
    -\rmi\Sigma^<_c(t_1,t_2) = \sum_{N=1} \sum_{a,b,I,P} (-1)^{|P|} \int_{\gamma_1} \rmd z_\Ical \int_{\gamma_2} \rmd z'_\Ical\, B^{(a)}_{N,I} (t_1,z_\Ical) \Upsilon^<_I(t_\Ical, t'_\Ical) B^{(b)*}_{N,P(I)} (t_2,z'_\Ical),
\end{equation}
with $\Upsilon^<$ as in \Eq{eq:big_G} (see \ref{app:prefactors}) where instead of integrations
over the branches $\gamma_-$ and $\gamma_+$ of equation (\ref{eq:sigma_time_ordered}) we now have integrations on the loops $\gamma_1$ and $\gamma_2$. 
In \Eq{eq:sigma_deformation_expansion} we can first focus our attention to the part integrated over $\gamma_1$, which by applying \Eq{eq:retarded_requirement} can be written as
\begin{equation}
    \int_{\gamma_1} \rmd z_\Ical\, B^{(a)}_{N,I} (t_1,z_\Ical) \Upsilon_I^<(t_\Ical, t'_\Ical) = \int \rmd t_\Ical \,
    B^{(a)R}_{N,I} (t_1,t_\Ical) \Upsilon_I^<(t_\Ical, t'_\Ical),
\end{equation}
with the shorthand notation $B^R(t_1,t_\Ical) = B^{R(1,\Ical)} (t_1,t_\Ical)$.
Here the $\Upsilon^<$ function can be left out of the retarded function since it does not depend on the branch indices for $z_\Ical$, but only on the real-times $t_\Ncal$ since all times on loop $\gamma_1$ are earlier than those on loop $\gamma_2$. This follows from the definition of the retarded component as a sum over the possible distributions of $z_\Ical$ between the two branches (see \ref{app:retarded_diagrams}), which leaves $\Upsilon^<$ the same in every term of the sum allowing it to be pulled out as a common factor.
After an analogous argument for the integrals over $\gamma_2$, \Eq{eq:sigma_deformation_expansion} takes the form \cite{Danielewicz1990}
\begin{equation} \label{eq:sigma_retarded}
    -\rmi\Sigma^<_c(t_1,t_2) = \sum_{N=1} \sum_{a,b,I,P} (-1)^{|P|} \int \rmd t_\Ical \rmd t'_\Ical\, B^{(a)R}_{N,I} (t_1,t_\Ical) \Upsilon_I^<(t_\Ical, t'_\Ical) B^{(b)R*}_{N,P(I)}  (t_2,t'_\Ical). 
\end{equation}
This equation is similar in form to \Eq{eq:sigma_time_ordered} and 
for the case of equilibrium systems it can analogously be Fourier transformed as
\begin{equation}  \label{eq:R-A_retarded} 
     - \rmi \Sigma_c^<(\omega)
    = \sum_{N=1} \sum_{a,b,I,P} (-1)^{|P|} \int \frac{\rmd\omega_\Ical}{(2\pi)^I} B^{(a)R}_{N,I} (\omega_\Ical) 2\pi \delta (\omega - \Omega) \Upsilon_I^<(\omega_\Ical) B^{(b)R*}_{N,P(I)} (\omega_\Ical).
 \end{equation}
Similar to the time-ordered case, equation \Eq{eq:R-A_retarded} can be derived for equilibrium systems without having to use the assumption of a continuous spectrum to argue that Matsubara integrals vanish from the expansion of $\Sigma_c^<$. This relies on an important result from
\ref{app:spectral_representation} where we derive spectral representations for general retarded and Matsubara diagrams, and show that these are directly related by analytical continuation. The crucial result is that for equilibrium systems there exists a general relation between the spectral form $\mathscr{D}^R (\omega_\Ncal)$ of a multi-retarded function describing a given Feynman diagram
and the spectral form $\mathscr{D}^M (\omega_N)$ of the corresponding multi-argument Matsubara function for the same Feynman diagram; this relation is given by
\begin{equation}
     \mathscr{D}^{R(1,2,\ldots,N)} (\omega_\Ncal )=
     \mathscr{D}^M(\omega_\Ncal - \mu - \rmi\eta_\Ncal)
     \label{eq:DRDM}
\end{equation}
where $\omega_\Ncal=\{ \omega_1,\ldots,\omega_N \}$ represents a set of $N$ frequencies
and $\eta_\Ncal$ is a set of positive infinitesimals.
The precise form of these functions is defined by equations  (\ref{eq:DR_separate}) and (\ref{Matsubara_transform})
in \ref{app:spectral_representation} which for our discussion here represent spectral forms of half-diagrams. 
A relation of the form (\ref{eq:DRDM}) was also derived by Baier and Ni{\'e}gawa \cite{Baier1994} 
but this required the implicit assumption of a continuous spectrum to make certain terms vanish. Our derivation in \ref{app:spectral_representation} does not require this assumption and is
instead directly based on fluctuation-dissipation relations for Matsubara functions; additionally it yields a set of useful Feynman rules to evaluate multi-retarded functions in frequency space.
Relation (\ref{eq:DRDM}) is
a generalization of \Eq{eq:S_M_S_R_relation} and can in particular be applied to the half-diagrams appearing in the retarded expansion
for the self-energy. 
In connection with step (iv) of Jeon's derivation \cite{Jeon1993} that we outlined in the previous section this 
allows us to convert half-diagrams appearing in the expansion of the Matsubara self-energy to
retarded half-diagrams which then precisely yields the retarded cutting rules that we derived above.
The final result is that, for the case of finite temperature equilibrium systems, 
equation \Eq{eq:R-A_retarded} is also valid without the assumption of a continuous spectrum, which means that in equilibrium the result holds also for finite systems.

Let us illustrate the Feynman rules derived in \ref{app:spectral_representation} for the evaluation of the retarded half-diagrams in expressions like \Eq{eq:R-A_retarded}. For example, in the cut diagram
\begin{equation} \label{eq:R_diagram_example}
    -\rmi D^<(\omega) = \bigdiagramcenter{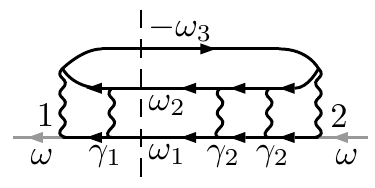} = \int \frac{\rmd\omega_\Ical}{(2\pi)^I} B^{(a)R}_{1,(123)}(\omega_\Ical) 2\pi \delta(\omega - \Omega) \Upsilon^<(\omega_\Ical) B^{(b)R*}_{1,(123)}(\omega_\Ical)
\end{equation}
where $\Omega = \omega_1 + \omega_2 + \omega_3$, the explicit form of $B^{(a)R}_{1,(123)}(\omega_\Ical)$ follows from
the rules given below \Eq{eq:time_ordered_equation} in \ref{app:spectral_representation}. We obtain
\begin{equation} \begin{split}
    &\left[ \bigdiagramcenter{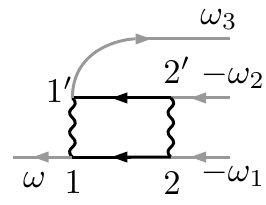} \right]^{R(1,1'22')} \\
    &= -2\pi \rmi \delta(\omega + \Omega) \int \frac{\rmd\nu \rmd\nu'}{4\pi^2} \frac{g^>(\nu) g^>(\nu')- g^<(\nu) g^<(\nu')}{\omega_1 + \omega_2 + \nu + \nu' - \rmi\eta} v_1 v_2 = D^{(a)R}(\omega, \omega_\Ical),
\end{split} \end{equation}
which is related to $B^{(a)R}_{1,3}(\omega_\Ical)$ (defined as in \Eq{eq:B_TI-t}) by
\begin{equation} \begin{split}
    D^{(a)R} (\omega, \omega_\Ical) &= \int \rmd t_1 \rmd t_\Ical\, B^{(a)R}_{1,(123)}(t_1, t_\Ical) \rme^{\rmi\omega t_1 + i\omega_\Ical \cdot t_\Ical} = 2\pi \delta(\omega + \Omega) B^{(a)R}_{1,(123)}(-\omega_\Ical) .
\end{split} \end{equation}
These expressions contain the interaction lines $v_1$ and $v_2$ 
as well as the functions $g^\lessgtr (\nu)$ which are the Fourier transforms of the functions $g^\lessgtr$ with respect to the difference of their time arguments,
where in all the expressions we suppressed spatial integrations for clarity of presentation.

We further note that it is always possible to expand a retarded half-diagram in terms of time-ordered Green's functions by splitting the loop integral into an integration on a forward and on a backward branch, but since we have two half-diagrams now we have two $\gamma_-$ branches and two $\gamma_+$ branches to keep track of.
Consider for example the second term in \Eq{eq:retarded_expansion}; expanding the half-diagrams leads to
\begin{equation} \label{eq:double_time_expansion}
    \bigdiagramcenter{T_matrix_retarded_2} = \bigdiagramcenter{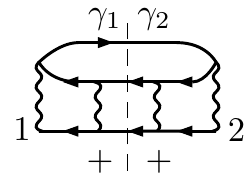} + \bigdiagramcenter{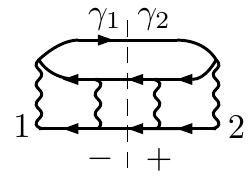} + \bigdiagramcenter{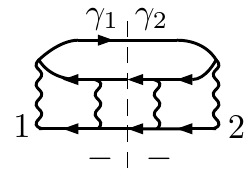} + \bigdiagramcenter{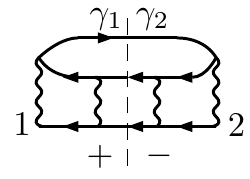}.
\end{equation}
The designation of the external vertices is irrelevant
since they are at end points of the loops in \Eq{eq:deformed_contour}.
Note that in evaluating these diagrams the Green's functions crossing the cut are independent of the $+$'s and the $-$'s, they are determined separately by the ordering of the loops. Alternatively, one could remove the dashed lines and use a different mark for each of the four branches, along with rules for the resulting $4^2=16$ Green's function components. An approach along these lines was developed by Bedaque, Das and Naik in \cite{Bedaque1997} starting from the latest time equation \cite{Veltman1963,Veltman1994} rather than referring to integration branches (see \cite{Gelis1997} for an extensive discussion). Either way, the time-ordered expansion leads to a very large number of diagrams, especially for higher order functions since each new external parameter comes with another forward and backward branch. It is therefore both more convenient, as well as physically more appealing, to make use of retarded half-diagrams; they have the advantage of giving only a single real-time diagram for each cut.

In summary: the use of retarded half-diagrams solves the crucial issue of non-vanishing vacuum diagrams while retaining the construction procedure of PSD approximations just as in the time-ordered formalism. It has the additional advantage that retarded half-diagrams have a physical interpretation as scattering amplitudes by past processes such that Hermitian products of retarded half-diagrams can be interpreted as scattering probabilities \cite{Danielewicz1990}. In the following chapter we will apply this approach to some commonly employed approximations to show that they are PSD.

\section{Application: positivity of diagrammatic approximations}

It is known (see e.g. section 9.7 in \cite{Stefanucci2013}) that equation (\ref{eq:Glessgtr}) is valid for
finite temperature systems in equilibrium or for non-equilibrium systems in a steady state 
limit when the Riemann-Lebesque lemma applies. 
It therefore follows directly that a PSD self-energy with a positive rate function $\Gamma(\omega)$ leads to a PSD Green's function with positive spectral function $A (\omega)$, and therefore it is sufficient to consider the PSD
approximations for the self-energy in order to obtain them for the spectral function.

In the following we will prove that some commonly used approximations for the self-energy, namely the second Born, T-matrix and GW approximations, preserve positivity also for finite temperature systems. The derivation illustrates the fact that the cutting rules derived for zero-temperature systems can be directly applied, provided we interpret the half-diagrams to be retarded rather than time-ordered functions. Consequently all methodology discussed in earlier work \cite{Stefanucci2014}, such as the construction of a minimal PSD-extension, apply directly to the finite temperature case as well.

\subsection{The T-matrix approximation}

The $T$-matrix approximation comes in two flavors, the $T$-matrix approximation in the particle-particle channel and the $T$-matrix approximation in the particle-hole channel.
The first one plays an important role in the study of systems with short range interactions such as Bose/Fermi gases in cold atom traps \cite{Shi1998_1,Perali2004,Chen2005_1} and nuclear matter \cite{Dickhoff2004,Alm1996,Bożek1999,Frick2003}, as well as in the description of Hubbard systems \cite{pavlyukh2021,PuigVonFriesen2010,Schlünzen2016}, to name a few examples. The second is important to describe
excitations in semi-conductors \cite{Kwong2009,strinati1988}, and in particular for the study of exciton states \cite{Piermarocchi2001}. The derivation below is given in the particle-particle channel but a completely analogous approach applies to the particle-hole channel. 

The particle-particle $T$-matrix approximation is given by
\begin{equation} \begin{split}
    \Sigma(1,2) &= \bigdiagramcenter{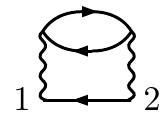} + \bigdiagramcenter{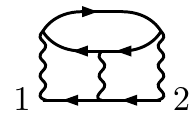} + \bigdiagramcenter{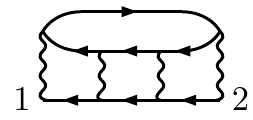} + \ldots \\
    &+ \bigdiagramcenter{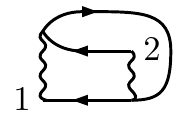} + \bigdiagramcenter{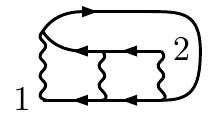} + \bigdiagramcenter{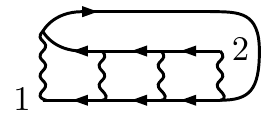} + \ldots,
\end{split} \end{equation}
where the first line contains the direct part and the second line the exchange part.
Performing the retarded expansion, and leaving out any diagrams with Matsubara integrals or disconnected half-diagrams, gives for the direct part
\begin{equation} \begin{split} \label{eq:T_matrix}
    -\rmi\Sigma^<_d(1,2) &= \bigdiagramcenter{T_matrix_1} + \bigdiagramcenter{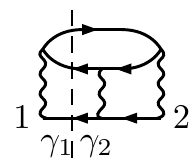} + \bigdiagramcenter{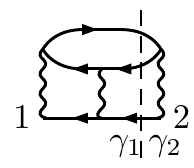} \\
    &+ \bigdiagramcenter{T_matrix_retarded_4} + \bigdiagramcenter{T_matrix_retarded_2} + \bigdiagramcenter{T_matrix_retarded_1} + \ldots \,.
\end{split} \end{equation}
Note that in the case of the $T$-matrix the ability to discard disconnected half-diagrams reduces the number of diagrams generated by $n$-th order contour diagram from $2^{n-2}$ in the time-ordered formalism to $n-1$ in the retarded formalism.

We factorize the Green's functions crossing the dashed line in \Eq{eq:T_matrix} using the relations \Eq{eq:tilde_less} and \Eq{eq:tilde_great} to obtain for the direct part
\begin{equation}
-\rmi\Sigma^<_d(1,2) = \left[ \bigdiagramcenter{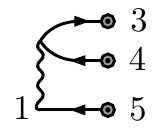} + \bigdiagramcenter{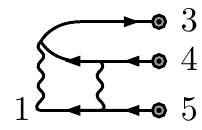} + \ldots \right] \left[ \bigdiagramcenter{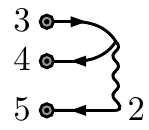} + \bigdiagramcenter{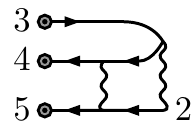} + \ldots \right].
\end{equation}
The expansion and factorization of the exchange part is analogous, expect that on the right hand half-diagram the labels $4$ and $5$ are exchanged. 
We therefore have
\begin{equation}
-\rmi\Sigma^<_x(1,2) = \left[ \bigdiagramcenter{T_matrix_half_L1} + \bigdiagramcenter{T_matrix_half_L2} + \ldots \right] \left[ \bigdiagramcenter{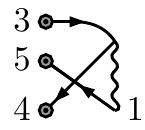} + \bigdiagramcenter{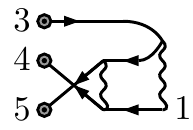} + \ldots \right]
\end{equation}
and the full self-energy can be written as
\begin{equation} \begin{split} \label{eq:T-matrix_half_diagrams}
&-\rmi\Sigma^<(1,2) \\
&= \left[ \bigdiagramcenter{T_matrix_half_L1} + \bigdiagramcenter{T_matrix_half_L2} + \ldots \right] \left[ \bigdiagramcenter{T_matrix_half_R1} + \bigdiagramcenter{T_matrix_half_R2} + \ldots + [345 \rightarrow 354] \right].
\end{split} \end{equation}
As discussed in \cite{Stefanucci2014}, this is PSD since the right-hand side is a sum 
of terms where the cut labels are permuted according to the subgroup $\{ \iota,(45) \}$
of the full permutation group. We finally remark this also proves that the second Born approximation is PSD since
it is the subset of the particle-particle $T$-matrix approximation corresponding to the second order diagrams.



\subsection{The GW approximation}

The $GW$ approximation is one of the most prevalent approximations in electronic structure theory \cite{Hedin1965,Aryasetiawan1998}. It
can be applied at different levels of self-consistency \cite{Stan2009}, leading 
to various flavors of the approximation designated as the $G_0 W_0$, $G W_0$ and the
fully self-consistent $GW$ approximation, which often are all grouped under the term $GW$ approximation. The physics of long range screening and plasmon dynamics that
is incorporated in this approximation plays an important role not only in electronic systems but also in charged plasmas at elevated temperatures \cite{Wierling1998,Fortmann2008}. 
An in-depth study of the finite temperature $GW_0$ approximation for high temperature plasmas was performed by Fortmann \cite{Fortmann2008} while a combination of the $GW$ approximation and cumulant expansion was recently studied by Kas and Rehr for warm dense matter\cite{Kas2017}. 
Furthermore, the finite temperature $GW$ approximation was recently studied for the calculation of the Helmholtz free energy and spin response functions \cite{Pokhilko2021}.

The $GW$ approximation for the self-energy, given by
\begin{equation}
    \Sigma_{GW}(1,2) = \bigdiagramcenter{T_matrix_1} + \bigdiagramcenter{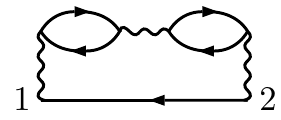} + \ldots,
\end{equation}
can be expressed in terms of retarded half-diagrams as
\begin{equation}
   -\rmi\Sigma^<_{GW}(1,2) = \left[ \bigdiagramcenter{T_matrix_half_L1} + \bigdiagramcenter{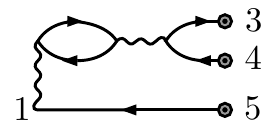} + \ldots \right] \left[ \bigdiagramcenter{T_matrix_half_R1} + \bigdiagramcenter{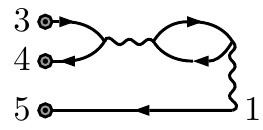} + \ldots \right],
\end{equation}
which shows that the expansion is positive. This result is in agreement with the numerical results of Fortmann \cite{Fortmann2008} for the $GW_0$ approximation.
The dressing of the base line of the self-energy diagram, that is required to deal with the $GW_0$ approximation, is readily included in our method since positivity is preserved during
the self-consistency iteration and the dressed line can therefore be cut as a Hermitian product. Similar arguments for the zero-temperature case were presented
in \cite{Stefanucci2014}. A completely analogous reasoning applies to the fully self-consistent $GW$ approximation.\\
We finally remark that both the GW and T-matrix approximations are not only PSD but also conserving \cite{Stefanucci2013}. However, a general conserving approximation will typically not be PSD as it need not have the form of a Hermitian product.
A minimal PSD extension of the approximation typically leads to the introduction of only specific cuts of additional higher order diagrams which render it PSD but not $\Phi$-derivable \cite{Stefanucci2013}.  
The situation at finite temperature is thus fully analogous to the one discussed in the zero-temperature case \cite{Stefanucci2014}.

\section{Conclusions}

In this work we addressed the question how to enforce important positivity constraints on spectral functions obtained from approximate diagrammatic expansions
in many-body theory of quantum systems at finite temperature.
We showed that this is possible using a finite temperature generalization of 
earlier work \cite{Stefanucci2014} that was based on cutting rules for Feynman diagrams and the construction of minimal PSD extensions. A straightforward expansion of that work, directly based on the Lehmann representation, was shown to be possible but turned out to be cumbersome for applications. An alternative formulation starting from self-energy diagrams, and the use of a factorization of Green's function lines along a cut, was found to be much more suitable. This lead to the consideration of half-diagrams with modified external legs which can be used to build finite temperature PSD approximations. However, when used in conjunction with the standard (anti)-time ordered formulation of many-body theory the PSD construction results in the appearance of vacuum diagram contributions which are absent in an exact expansion and which therefore should not contribute in any approximate expansion. This issue was finally resolved by a reformulation based on a deformation of the integration contour, which leads to an expansion in retarded half-diagrams for which such vacuum diagram contributions automatically vanish.
Apart from providing an attractive physical interpretation the formalism also extends to non-equilibrium systems in initial equilibrium and thus allows us to study physical processes such as the steady-state limit of quantum transport situations, thereby generalizing earlier work for zero-temperature systems \cite{Hyrkas2019b}. Our derivations furthermore provide a unified view of various earlier studies that addressed cutting rules in finite temperature systems and which can be related to each other by means of contour deformations.
Additionally we derived a generalized connection between retarded and Matsubara multi-argument functions and provided a useful set of Feynman rules for multi-retarded functions in frequency space. 
Finally we established that important commonly used approximations, namely the $GW$, second Born and $T$-matrix approximation, retain positive spectral functions at finite temperature. The work opens up new fields for application of finite temperature many-body theory, notably in studying vertex corrections and higher order processes in finite temperature many-particle systems as was done in the zero-temperature version of the theory \cite{Pavlyukh2016,Pavlyukh2020}.

\ack
D.K. acknowledges the academy of Finland for funding under Project No. 308697. M.H. thanks the Finnish Cultural Foundation for support. R.v.L. acknowledges the academy of Finland for funding under Project No. 317139.

\appendix

\section{Positive semi-definite Hermitian forms}
\label{app:PSDHF}

In this appendix we will show that the (weighted) Hilbert-Schmidt product
\begin{equation} \label{eq:HSP}
 \dlangle \Ah | \Bh \drangle = \trace{\rhoh \Ah^\dagger \Bh},
\end{equation}
satisfies the properties of a positive semi-definite Hermitian form and the Cauchy-Schwartz inequality.
It is readily seen that \Eq{eq:HSP}  satisfies the properties of a Hermitian form. We have
\begin{align}
\dlangle \hat{A} | \lambda_1 \hat{B}_1 +   \lambda_2 \hat{B}_2 \drangle &=  \lambda_1   \dlangle \hat{A} | \hat{B}_1 \drangle +
 \lambda_2   \dlangle \hat{A} | \hat{B}_2 \drangle \\
  \dlangle \Ah | \Bh \drangle &=  \dlangle \Bh | \Ah \drangle^* 
\end{align}
where $\lambda_1$ and $\lambda_2$ are complex numbers. 
We further have that
\begin{equation}
\dlangle \Ah | \Ah \drangle  = \trace{ \hat{\rho} \, \hat{A}^\dagger \hat{A}} \geq 0
\label{eq:seminorm2}
\end{equation}
which makes our Hermitian form positive semi-definite. This is readily derived as follows.
Since $\rhoh$ is a PSD operator, it can be uniquely written as $\hat{\rho} = \hat{\rho}^{\frac{1}{2}} \hat{\rho}^{\frac{1}{2}}$. Furthermore, since $\rhoh$ is self-adjoint, $\hat{\rho}^{\frac{1}{2}}$ is self-adjoint as well. Using the cyclic property of the trace we can write
\begin{equation}\label{eq:operatorProductForm}
 \dlangle \Ah | \Ah \drangle  = \trace{ \rhoh^{1/2} \Ah^\dagger \Ah \rhoh^{1/2} } = \trace{\Fh^\dagger \Fh},
\end{equation}
with $\Fh = \Ah \rhoh^{1/2}$. Since any operator of the form $\Fh^\dagger \Fh$ is PSD, the trace is non-negative, which means that $\dlangle \Ah | \Ah \drangle \geq 0$. 
To make the Hilbert-Schmidt product a proper inner product we also need the property
\begin{equation}
 \dlangle \Ah | \Ah \drangle  = 0 \quad \Rightarrow \quad \Ah =0.
 \label{property}
\end{equation}
However, from \Eq{eq:operatorProductForm} we can only prove that
\begin{equation}
\dlangle \Ah | \Ah \drangle  = 0 \quad \Rightarrow \quad \Ah  \rhoh^{1/2}=0.
\end{equation}
If $\rhoh$ is a strictly positive operator this implies $\Ah=0$ and we are indeed dealing with a proper inner product, as was proven by Haag et al \cite{Haag1967}.
This is, for example, the case if $\rhoh$ represents a grand canonical ensemble. However, if $\rhoh$ is only a positive semi-definite operator then we do not in general
have an inner product but instead what is known in the mathematical literature as a positive semi-definite Hermitian form (PSDHF). This structure is sufficient to derive the Cauchy-Schwartz inequality
\begin{equation}\label{eq:cauchySchwarz}
 |\dlangle \Ah | \Bh \drangle|^2
 \leq
 \dlangle \Ah | \Ah \drangle \dlangle \Bh | \Bh \drangle.
\end{equation}
Equation \Eq{eq:cauchySchwarz} is obviously true if $\dlangle \Ah  | \Bh  \drangle =0$
so it is sufficient in the following to assume that $\dlangle \Ah  | \Bh  \drangle \neq 0$.
We use property (\ref{eq:seminorm2}) and consider
\begin{equation}
0 \leq \dlangle \Ah - \lambda \Bh | \Ah - \lambda \Bh \drangle  = \dlangle \Ah  | \Ah  \drangle - \lambda  \dlangle \Ah  | \Bh  \drangle -
\lambda^*  \dlangle \Bh  | \Ah  \drangle  + |\lambda|^2 \dlangle \Bh  | \Bh  \drangle
\label{inequality1}
\end{equation}
where $\lambda$ is a complex number that we can choose freely.
Let us now write $\dlangle \Ah  | \Bh  \drangle = | \dlangle \Ah  | \Bh  \drangle | \rme^{\rmi \phi}$ and choose $\lambda =  s\, \rme^{-\rmi\phi}$ where $s$ is real.
Then \Eq{inequality1} becomes
\begin{equation}
0 \leq  \dlangle \Ah  | \Ah  \drangle - 2 s \, | \dlangle \Ah  | \Bh  \drangle |  + s^2 \, \dlangle \Bh  | \Bh  \drangle.
\label{inequality2}
\end{equation}
Since $\dlangle \Ah  | \Bh  \drangle \neq 0$ it follows immediately that also $\dlangle \Bh  | \Bh  \drangle \neq 0$ since otherwise \Eq{inequality2}
would represent a non-constant linear function in $s$ bounded from below which does not exist. The quadratic polynomial in $s$ on the right hand side of the inequality in 
\Eq{inequality2} is therefore a positive semi-definite parabola-shaped function which can at most have one real zero. It therefore follows that its corresponding discriminant is zero or negative, i.e.
\begin{equation}
|\dlangle \Ah | \Bh \drangle|^2
 -
 \dlangle \Ah | \Ah \drangle \dlangle \Bh | \Bh \drangle \leq 0
\end{equation}
which is equivalent to \Eq{eq:cauchySchwarz}.

The Cauchy-Schwartz inequality can be used to obtain bounds on various correlators. 
As an example, let us consider the lesser Green's function given by
\begin{equation}\label{eq:gLesser}
 -\rmi G^<(1,2) = \trace{\rhoh\, \psihd_H (2) \psih_H(1)} =  \dlangle \psih_H (2) | \psih_H (1) \drangle,
\end{equation}
with the shorthand notation $1 = \xb_1 t_1$ etc.
Applying the Cauchy-Schwartz inequality to \Eq{eq:gLesser} yields the upper bound
\begin{equation}\label{eq:upperBoundGl}
 |G^<(1,2)|^2 \leq  n(1) n(2) .
\end{equation}
As such, the magnitude of the lesser Green's function is bounded from above by the density in a general time-dependent system. In systems where the density decays exponentially for large distances, such as in atoms and molecules, \Eq{eq:upperBoundGl} states that also the lesser Green's function decays exponentially (or faster). The upper bound is more familiar for the time diagonal,  $t_1=t_2$, where \Eq{eq:upperBoundGl} yields a well-known \cite{Kutzelnigg2007} upper bound for the time-dependent single-particle density matrix.


\section{Feynman rules for half-diagrams}
\label{app:prefactors}

The main goal of this appendix is to show that \Eq{eq:AAP} is valid provided that
we assign a proper topological pre-factor to each half-diagram. 
We are concerned about this since
each half-diagram $A_{N,I}^{(a)}$ 
contains modified external legs $\tilde{g}^{\lessgtr}$ 
which replaced the standard Green's functions $g$
and therefore is not obtained directly from an expansion involving Wick's theorem, as was the case for the Lehmann amplitudes in the zero-temperature formulation. 
We thus need to assure that 
we can assign pre-factors to the half-diagrams such that
the gluing operation leads to the correct prefactors for the self-energy.
In order to clearly distinguish the external vertices of the self-energy from the labels of the cut lines we will label
them $\alpha$ and $\beta$ instead of $1$ and $2$, i.e. we consider the self-energy $-i \Sigma^< (\alpha,\beta)$.

The pre-factor that we will use for the half-diagrams is determined by the following rule.
For a given half-diagram $A_{N,I}^{(a)}(\alpha)$ attached to external self-energy vertex $\alpha$ we denote the entrance lines by $j'$ and the exit ones by $j$.
Since $\alpha$ corresponds to an exit vertex there are $N$ labels $j$ and $N+1$ labels $j'$.
For any such labeling the half-diagram is interpreted as belonging to an expansion of $G_{N+1} (\alpha,J;J')$ 
where $J$ and $J'$ are the sets of labels for $j$ and $j'$. 
We then assign to the half-diagram the topological pre-factor \cite{negele1995}
\begin{equation}
\rmi^{n_v} (-1)^{N+1 + l + \tilde{l}}
\label{eq:prefactor}
\end{equation}
where $n_v$ is the number of $v$-lines, $l$ the number of loops formed by the $g$-lines and 
$\tilde{l}$ the number of extra loops created when we imagine each 
entrance vertex from the set $J'$ to be merged with the exit vertex from $(\alpha,J)$
with the same position in the argument list.
It follows from our definition that the topological pre-factor for the diagram $A_{N,P(I)}^{(a)}(\alpha)$ changes
with a relative factor $(-1)^{|P|}$ compared to that of $A_{N,I}^{(a)}(\alpha)$ since every interchange of external vertices changes the number of loops for $\tilde{l}$ by one. Note, however, that the actual value of the permuted diagram does not just differ by a sign from the non-permuted one as it represents a topologically different diagram.
The central statement that we will demonstrate is that (\ref{eq:AAP})
gives the correct expansion for the self-energy, i.e. with the proper topological pre-factor of the Feynman rules for the self-energy. To show this we need two ingredients; the first one (proven in \ref{app:retarded_diagrams}) is that
\begin{equation}
    A_{N,I}^{(a)\dagger}(\alpha) =(-1)^{N} \, \tilde{A}_{N,I}^{(a)} (\alpha)
    \label{eq:A_conj}
\end{equation}
where we defined $\tilde{A}_{N,I}^{(a)}$ to be the non-conjugated half-diagram but with all directions of the Green's function lines reversed and the $\tilde{g}$ legs replaced by $\tilde{g}^\dagger$ and vice versa. 
Such a reversed diagram represents the right-hand side of a cut self-energy diagram.

The second ingredient is a topological rule that we state as follows.
Let
$l(ab)$ be the number of loops created when two half diagrams of topology $a$ and $b$ are glued together, then
\begin{equation}\label{eq:prefactor_relation}
  (-1)^{\tilde{l}{(a)} + \tilde{l}{(b)}} = (-1)^{N+l{(ab)}}
\end{equation}
where $\tilde{l} (a)$ and $\tilde{l}(b)$ are the factors from our rule stated above for the two separate half-diagrams. We postpone the demonstration of the latter statement and first establish that it implies the desired result
in combination with (\ref{eq:A_conj}). We consider the product
\begin{equation}
    \sum_{I} (-1)^{|P|}A_{N,I}^{(a)} (\alpha) A_{N,P(I)}^{(b)\dagger} (\beta) ,
    \label{eq:permute_A}
\end{equation}
which amounts to a gluing operation.
As we outlined above, for the purpose of determining the topological pre-factor it is sufficient to consider
the particular product
\begin{equation}
    \sum_{I} A_{N,I}^{(a)} (\alpha) A_{N,I}^{(b)\dagger} (\beta)
    \label{eq:permute_A2}
\end{equation}
since permutation of the external vertices changes the topological prefactor by $(-1)^{|P|}$.
From (\ref{eq:A_conj}) it follows that conjugation gives a diagram with reversed Green function lines which is accompanied with a pre-factor $(-1)^N$ while the gluing of the $\tilde{g}$ lines using \Eq{eq:tilde_less} and \Eq{eq:tilde_great} produces an overall factor of $-\rmi$ as we glue one more hole line than a particle line. The gluing procedure thus yields a net factor of
$-\rmi (-1)^N$ and the overall factor of the diagram is therefore
\begin{equation}
    C= -\rmi (-1)^N \rmi^{n_v} (-1)^{l(a)+ l(b) + \tilde{l} (a) + \tilde{l} (b)} 
\end{equation}
where $n_v=n_v (a) + n_v (b)$ is the total number of interaction lines. Then equation (\ref{eq:prefactor_relation}) immediately gives that
\begin{equation}
    C= -\rmi (-\rmi)^{n_v} (-1)^l
\end{equation}
where $l$ is the total number of loops in the glued self-energy diagram. Given that we were considering
a diagram for $-\rmi \Sigma^<$ we see that we precisely recover the usual Feynman rule prefactor 
$(-\rmi)^{n_v} (-1)^l$ for the self-energy diagram, which thereby proves the desired relation (\ref{eq:AAP}).

To prove (\ref{eq:prefactor_relation})
we consider
the factorization of a general self-energy diagram in terms of half-diagrams from \Eq{eq:permute_A2}.
For the left hand diagram we 
label the entrance vertices of the cut lines with labels $i'$. 
If a path of directed Green's function lines starting at entrance vertex $i'$ exits the half-diagram at a given vertex we label that exit vertex by $i$.
This ensures that connecting $i$ to $i'$ in the left-hand diagram always creates one loop, and therefore $(-1)^{\tilde{l}{(a)}} = (-1)^{N+1}$. 
For example: 
\begin{equation} \label{eq:labeling_example}
    \bigdiagramcenter{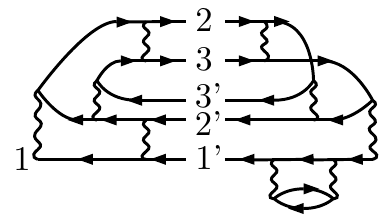} \xrightarrow{} \bigdiagramcenter{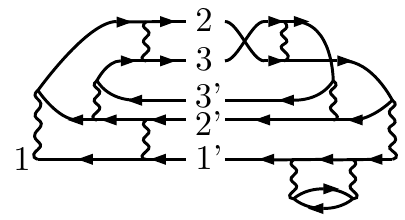}.
\end{equation}
The right hand side can always be obtained by a permutation $Q$ of the external vertices from a
diagram with $\tilde{l}(b)=N+1$ loops (as shown graphically in \Eq{eq:labeling_example}) which yields a factor $(-1)^{\tilde{l}(b)}=(-1)^{N+1+ |Q|}$.
Every permutation that changes $\tilde{l} (b)$ by one also changes $l(ab)$ by one and
therefore $(-1)^{l(ab)} = (-1)^{N + |Q|}$ since we do not glue the entrance and exit vertices of the self-energy diagram. We thus obtain $(-1)^{\tilde{l}{(a)} + \tilde{l}{(b)}} = (-1)^{N+l{(ab)}}$ as was to be proven.

\section{Multi-retarded functions and their adjoints}
\label{app:retarded_diagrams}

The multi-retarded $N$-point function occurring in equation \Eq{eq:retarded_requirement} is defined as (see \cite{Danielewicz1990, Hyrkas2019})
\begin{equation} \label{eq:retarded_definition}
D^{R(1,2 \cdots N)}(t_\Ncal) = \sum_P \theta(t_1, t_{P(2)},\ldots,t_{P(N)}) D^{[1,P(2),\ldots,P(N)]}(t_\Ncal)
\end{equation}
where $t_\Ncal=(t_1,\ldots, t_N)$
and $\theta (t_1,\ldots,t_N)$ is defined to be equal to one when $t_1 > \ldots > t_N$ and zero otherwise.
Since $t_1$ is always the latest time such a function is called a retarded function with respect to top element $1$.
Here the sum is over all permutations $P$ of $2,\ldots,N$ and $[1,2,\ldots,N] = [[[1,2],3],\ldots,N]$ is a nested commutator 
represented by a formal sum of strings of integers such as $[1,2]=12-21$. In general the commutator is of the form
$L=\sum_i \sigma_i l_i$ where $\sigma_i=\pm 1$ and $l_i$ is a string of $N$ integers and we define
\begin{equation}
    D^L = \sum_i \sigma_i D^{l_i}
\end{equation}
where $D^{l_i}$ corresponds to a real time function for the contour ordering $l_i$.
For example for $N = 3$ we would have
\begin{equation}
D^{R(1,23)}(t_1,t_2,t_3) = \theta(t_1,t_2,t_3) D^{[1,2,3]}(t_1,t_2,t_3)  + \theta(t_1,t_3,t_2) D^{[1,3,2]}(t_1,t_2,t_3),
\end{equation}
where
\begin{equation}
D^{[1,2,3]} = D^{[[1,2],3]} = D^{123 - 312 - 213 + 321} = D^{123} - D^{312} - D^{213} + D^{321}.
\end{equation}
The string of indices in the superscript gives the contour order of the time arguments with the latest argument to the left.
For example, $D^{312}$ corresponds to a real-time function for the contour ordering $z_3 > z_1 > z_2$. \\
The aim is to derive a useful expression for the adjoint of the multi-retarded function $D^R$
\begin{equation} \begin{split}
\left[ D^{R(1,2\cdots N)}(t_\Ncal) \right]^\dagger &= \sum_P \theta(t_1, t_{P(2)},\ldots,t_{P(N)}) \left[ D^{ [1,P(2),\ldots,P(N)]}(t_\Ncal) \right]^\dagger,
\label{eq:DR_adjoint}
\end{split} \end{equation}
which requires us to consider the adjoint of its various components $D^{J}$
where $J=j_1 \ldots j_N$ is a reordering of the labels $1\ldots N$.
Let us take $D_n^{J} (t_\Ncal)$ (where we added a sub-index $n$) to be a specific ordered component of an un-integrated Feynman diagram for a many-particle Green's function $G_n$. By un-integrated diagram we mean that we apply all Feynman rules to it apart from integrating over internal times; the diagram therefore depends on $N=2n+n_v$ times where $n_v$ is the number of interaction lines in $D^J_n$.
By taking the adjoint of such a component
every internal $g$-line is transformed to 
\begin{equation} \label{eq:g_con}
    g^{\lessgtr \dagger}(1,2) = -g^\lessgtr(2,1),
\end{equation}
which thus reverses the direction of each $g$-line, thereby reserving the time-ordering of $D_n^J$
and since the diagram prefactor contains an $\rmi$ for each interaction line, we get a minus sign for each interaction as well. The adjoint for an ordered diagram with $n_g$ Green's function lines and $n_v$ interaction lines is therefore given by
\begin{equation} \label{eq:ordered_conjugation}
 D_n^{J \dagger} (t_\Ncal)  = (-1)^{n_g + n_v} \tilde{D}_n^{\bar{J}}(t_\Ncal) = (-1)^{N-n} \tilde{D}_n^{\bar{J}}(t_\Ncal)
\end{equation}
where $\tilde{D}^J_n$ is the diagram obtained from $D_n^J$ by reversal of all the $g$-lines
and $\bar{J}$ is string $J$ in reverse order. We further used that $n_g=2n_v+n$ and $N=n_v+2n$ to rewrite the prefactor in the second step of equation (\ref{eq:ordered_conjugation}).
We now employ this relation in the expression (\ref{eq:DR_adjoint}).
If a nested commutator $[1,\ldots,N]= \sum \sigma_i l_i$ contains  a specific ordering of $l_i$, it also contains the reverse ordering $\bar{l}_i$, with a relative $+/-$ sign if $N$ is odd/even; consequently reversing the ordering in every string $l_i$
yields an overall factor $(-1)^{N-1}$. Together with equation (\ref{eq:ordered_conjugation}) we thus obtain
\begin{equation} \label{eq:R_conjugation}
\left[ D_n^{R(1,2\cdots N)}(t_\Ncal) \right]^\dagger = (-1)^{n - 1} \tilde{D}_n^{R(1,2\cdots N)}(t_\Ncal).
\end{equation}
This result can now be employed to deduce similar relations for
the half-diagrams $B^{(a)R}_{N,I}(t_1,t_\Ical)$ appearing in the retarded expansion of self-energy diagrams \Eq{eq:sigma_retarded}. They are (now integrated) $G_{N+1}$ diagrams with their external legs amputated, but still follow the relation \Eq{eq:R_conjugation}
as can be checked by following the same steps as outlined above. Thus we have for these half-diagrams 
\begin{equation}
    [B^{(a)R}_{N,I}(t_1,t_\Ical)]^\dagger = (-1)^N \tilde{B}^{(a)R}_{N,I}(t_1,t_\Ical).
    \label{eq:B_adjoint}
\end{equation}
The half-diagrams $A^{(a)R}_{N,I}$ are obtained from $B^{(a)R}_{N,I}$ by attaching $\tilde{g}^\lessgtr$-functions as external legs for the cut vertices $I$; from (\ref{eq:B_adjoint}) it then follows that
\begin{equation}
    [A^{(a)R}_{N,I}(t_1,t_0)]^\dagger = (-1)^N \tilde{A}^{(a)R}_{N,I}(t_1,t_0),
\end{equation}
when we define $\tilde{A}^{(a)R}_{N,I}$ as the diagram with the $\tilde{g}^\lessgtr$ legs replaced by $[\tilde{g}^\lessgtr]^\dagger$ and vice versa.
As a final remark we note that a similar relation can be derived for (anti)-time-ordered diagrams.
Since $(g_{--})^\dagger = -g_{++}$ this would have a given a factor $(-1)^{N+1}$ in the equation above. 
The gluing rule for the (anti)-time ordered functions (\ref{eq:g_T_cutting}) then generates an additional minus sign yielding a correct pre-factor
for every diagram in the expansion of the self-energy from the considerations in \ref{app:prefactors}.

\section{Analytic continuation and Feynman rules for multi-argument retarded and Matsubara functions}
\label{app:spectral_representation}

The aim of this appendix is to derive the important relation (\ref{eq:DRDM}) and to
give the Feynman rules for the evaluation of retarded half-diagrams in frequency space. 
To obtain a spectral representation of a multi-retarded function we define the Fourier transform
\begin{equation} \label{eq:fourier_transform}
    \mathcal{D}^{R(e,\Ical)}(\omega_\Ncal) = \int \rmd t_\Ncal\, \rme^{\rmi \omega_\Ncal \cdot t_\Ncal} D^{R(e,\Ical)}(t_\Ncal)
\end{equation}
where $\Ncal = \{ 1 , \ldots , N \}$ is an ordered set of labels and $t_\Ncal$ and $\omega_\Ncal$ are sets of arguments labeled with $\Ncal$, $e$ is the label for the external time and $\Ical = \Ncal \setminus e$ is the set $\Ncal$ with label $e$ removed. 
We further used the shorthand notation
\begin{equation}
\omega_\Ncal \cdot t_\Ncal = \omega_1 t_1 + \omega_2 t_2 + \ldots + \omega_N t_N.
\end{equation}
 When $D^R$ represents the integrand of a half-diagram we can set some of $\omega_j$'s to zero to recover the Fourier representation of any desired half-diagram discussed in the main text.
To carry out the Fourier integral it will be useful to use the relation (see \cite{Hyrkas2019} for more details)
\begin{equation} \label{eq:app5_start}
 \int_{\gamma} \rmd z_{\Ncal \setminus e}\, D(z_\Ncal) = \int_{t_0}^\infty \rmd t_{\Ncal \setminus e}\, D^{R(e,\Ical)}(t_\Ncal),
\end{equation}
where $D(z_\Ncal)$ is a contour function that can explicitly be written as a sum over contour-orderings as
\begin{equation} \label{eq:keldysh_function}
    D(z_\Ncal) = \sum_{P} \theta(z_{P(\Ncal)}) D^{P(\Ncal)}(t_\Ncal),
\end{equation}
where the sum is over all permutations $P(\Ncal)$ of the ordered set $\Ncal$.
The contour dependence only occurs in the contour Heaviside function $\theta (z_\Ncal)$ defined to be one for $z_1 > ...> z_N$ on the contour and zero otherwise. 
We take $D^{P(\Ncal)}(t_\Ncal)$ in \Eq{eq:keldysh_function} to have the structure
\begin{equation}
    D^{P(\Ncal)}(t_\Ncal) = \left[ C \prod_L g_L(z_{L_f}, z_{L_i}) \prod_I v_I \right]^{P(\Ncal)} ,
    \label{structure}
\end{equation}
where the term on the right hand side is a product of Green's functions $g_L$ and interaction lines $v_I$ and the superscript $P(\Ncal)$
denotes the ordering of the contour times used to evaluate this product. Hereby the Green's functions become greater and lesser functions and consequently the resulting expression is a function of real times only.
We furthermore denoted by $C$ the topological prefactor given by the Feynman rules for this diagram, and $L$ runs over all $g$-lines in the diagram starting from initial time $z_{L_i}$ and ending at final time $z_{L_f}$.
Position-spin arguments are here suppressed and the interaction lines are taken to be time-independent. A half-diagram can always
be assumed to be in this form for time-local interactions $v(z_1,z_2) = v\, \delta(z_1,z_2)$ as
the delta-functions can already be integrated out in the original self-energy diagram that gave rise to the half-diagram under consideration.

We express the permuted set as $P(\Ncal) = \{ \Ncal^P_+, e,  \Ncal^P_- \}$,
where $\Ncal^P_+$/$\Ncal^P_-$ are ordered sets of labels ordered after/before $e$. The contour-times $z_{\Ncal^P_+}$/$z_{\Ncal^P_-}$ are therefore placed on the backward/forward branch. This corresponds to
\begin{equation}
    \bigdiagramcenter{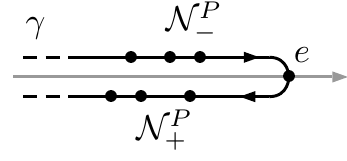}
\end{equation}
Taking an integral over the internal times for \Eq{eq:keldysh_function} then involves integration over the step-functions which can be expressed as
\begin{equation} \begin{split} \label{eq:step_function_split}
   \int_\gamma \rmd z_{\Ncal \setminus e}\, \theta(z_{P(\Ncal)}) &= \int_{\gamma_-} \rmd z_{\Ncal^P_-} \int_{\gamma_+} \rmd z_{\Ncal^P_+}\, \theta(t_e, t_{\bar{\Ncal}^P_+}) \theta(t_e, t_{\Ncal^P_-}) \\
   &= \int_{-\infty}^\infty \rmd t_{\Ncal \setminus e}\, (-1)^{N^P_+} \theta(t_e, t_{\bar{\Ncal}^P_+}) \theta(t_e, t_{\Ncal^P_-}),
\end{split} \end{equation}
where $\bar{\Ncal}$ is a set of reversed order and $(-1)^{N^P_+} = (-1)^{|\Ncal^P_+|}$ comes from the direction of integration on the backward branch \cite{Hyrkas2019}. Note that the contour $\gamma$ is here taken to extend to $-\infty$, and can be taken to extend to $+\infty$ since the integral is cut off by the step-functions at $t_e$. We can thus express the retarded component as
\begin{equation} \label{eq:multi_retarded}
    D^{R(e,\Ical)}(t_\Ncal) = \sum_P (-1)^{N^P_+} \theta(t_e, t_{\bar{\Ncal}^P_+}) \theta(t_e, t_{\Ncal^P_-}) D^{P(\Ncal)}(t_\Ncal).
\end{equation}
This relation can now be inserted into (\ref{eq:fourier_transform}).
Each $g$-line in the term (\ref{structure}) corresponds to either a lesser or a greater component which we rewrite in frequency space as
\begin{equation}
    g_L^\lessgtr(t,t') = \int \frac{\rmd\nu}{2\pi} g_L^\lessgtr(\nu) \rme^{-\rmi\nu(t-t')}
\end{equation}
and thus obtain
\begin{equation} \begin{split} \label{eq:phase_factors}
D^{P(\Ncal)}(t_\Ncal) &= \left( \prod_L \int \frac{\rmd\nu_L}{2\pi} \rme^{-\rmi\nu_L (t_{L_f} - t_{L_i})} \right) \left[ C \prod_L g_L(\nu_L) \prod_I v_I \right]^{P(\Ncal)}  \\
&= \int \frac{\rmd\nu_\Lcal}{2\pi^{|\Lcal|}} \rme^{-\rmi\nu_\Lcal \cdot (t_{\Lcal_f} - t_{\Lcal_i})} \mathcal{D}^{P(\Ncal)}(\nu_\Lcal),
\end{split} \end{equation}
where $|\Lcal|$ is the number of $g$-lines in the diagram and $\nu_\mathcal{L}$ is the set of the corresponding frequencies $\nu_L$.
This allow us to rewrite \Eq{eq:fourier_transform} as
\begin{equation} \begin{split} \label{eq:total_component}
    \mathcal{D}^{R(e,\Ical)}(\omega_\Ncal)
    = \sum_P 
    &\int \frac{\rmd\nu_\Lcal}{(2\pi)^{|\Lcal|}} \Lambda_{e P(\Ncal)}(\sigma_\Ncal) \mathcal{D}^{P(\Ncal)}(\nu_\Lcal),
\end{split} \end{equation}
where
\begin{equation} \label{eq:lambda}
   \Lambda_{eP(\Ncal)}(\sigma_{\Ncal}) = (-1)^{N^P_+} \int \rmd t_\Ncal\, \theta(t_e,t_{\bar{\Ncal}^P_+}) \theta(t_e, t_{\Ncal^P_-}) \rme^{\rmi \sigma_\Ncal \cdot t_\Ncal}.
\end{equation}
Here $\sigma_\Ncal = \{ \sigma_1,\ldots, \sigma_N \}$ where $\sigma_j$ is to the total energy leaving the vertex/vertices at time $t_j$ (in the example diagram given here black lines belong to the diagram and grey lines are potential external connections)
\begin{equation}
\sigma_j = \omega_j + \sum_{L,L_i = j} \nu_L - \sum_{L,L_f = j} \nu_L \qquad \bigdiagramcenter{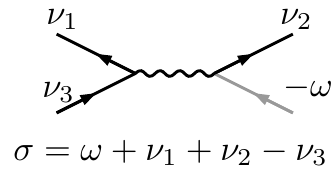}
\label{eq:sigma_sum_R}
\end{equation}
Here $\omega_j$ is the energy leaving the diagram at interaction line $j$, and therefore 
in a Feynman diagram we label the frequency on an incoming external line with $-\omega_j$ as
in the figure of (\ref{eq:sigma_sum_R}). The first sum adds the energy being carried away by internal $g$-lines, and the second sum subtracts the energy being brought in by internal $g$-lines.
The integrals in \Eq{eq:lambda} can be evaluated using the expression
\begin{equation} \begin{split} \label{eq:integral_result}
&\int \rmd t_1 \cdots \rmd t_k\, \theta(t, t_k, \ldots, t_1) \rme^{\rmi\sigma_1 t_1 + \ldots +\rmi\sigma_k t_k} \\
&= (-\rmi)^k \frac{\rme^{\rmi(\sigma_1 + \ldots + \sigma_k)t}}{(\sigma_1 - \rmi \eta_1)(\sigma_1 + \sigma_2 - \rmi \eta_2)\cdots (\sigma_1 + \ldots + \sigma_k - \rmi \eta_k)}
\end{split} \end{equation}
which can be derived by writing each step-function as
\begin{equation}
\theta(t, t_k, \ldots, t_1) = \theta(t - t_k) \theta(t_k - t_{k-1}) \cdots \theta(t_2 - t_1),
\end{equation}
and substituting the representation
\begin{equation}
\theta(t) = - \lim_{\eta \rightarrow 0} \frac{1}{2\pi i} \int_{-\infty}^\infty \rmd\xi \frac{\rme^{-\rmi\xi t}}{\xi + \rmi \eta}.
\end{equation}
After using equation (\ref{eq:integral_result}) in equation (\ref{eq:lambda}) 
we further use
\begin{equation}
    \int \rmd t_e \rme^{\rmi (\sigma_1 + \ldots + \sigma_N) t_e} =  2\pi \delta(\sigma_1 + \ldots + \sigma_N) =  2\pi \delta(\omega),
\end{equation}
where $\omega = \sum_i \omega_i=\sum_i \sigma_i$ since all internal energy flows cancel by summing over all vertices.
Defining a sum over the $l$ first or the $l$ last elements in $\sigma_\Ncal$
\begin{equation} \label{eq:Omega_def}
    \Omega^{\Ncal}_l = \sum_{i = 1}^l \sigma_i , \qquad \Omega^{\bar{\Ncal}}_l = \sum_{i = 1}^l \sigma_{N+1-l}
\end{equation}
then leads to
\begin{equation} \label{eq:L_def_1}
    \Lambda_{eP(\Ncal)}(\sigma_\Ncal) = \frac{(-1)^{N^P_+} (-\rmi)^{N-1} \,2\pi \delta(\omega)}{\prod_{k=1}^{N^P_-} (\Omega^{\overline{P(\Ncal)}}_k - \rmi\eta_k) \prod_{l=1}^{N^P_+} (\Omega^{P(\Ncal)}_l - \rmi\eta_l) }.
\end{equation}
A multi-retarded diagram \Eq{eq:total_component} therefore has the spectral representation
\begin{equation}
    \mathcal{D}^{R(e,\Ical)}(\omega_\Ncal) = 2\pi \delta(\omega) \mathscr{D}^{R(e,\Ical)}(\omega_\Ncal)
    \label{eq:DR_separate}
    \end{equation}
where we defined the analytic function
\begin{equation} \label{eq:retarded_equation}
    \mathscr{D}^{R(e,\Ical)}(\omega_\Ncal) = \sum_P \int \frac{\rmd\nu_\Lcal}{(2\pi)^{|L|}} \frac{(-1)^{N^P_+} (-\rmi)^{N-1} }{\prod_{k=1}^{N^P_-} (\Omega^{\overline{P(\Ncal)}}_k - \rmi\eta_k) \prod_{l=1}^{N^P_+} (\Omega^{P(\Ncal)}_l - \rmi\eta_l)} \mathcal{D}^{P(\Ncal)}(\nu_\Lcal).
\end{equation}
We can now also easily work out the spectral representation for a time-ordered diagram. The lack of the backward branch means that $N^P_+ = 0$, which leads to
\begin{equation} \begin{split} \label{eq:time_ordered_equation}
    \mathcal{D}^T(\omega_\Ncal) &= 2\pi \delta(\omega) \sum_P \int \frac{\rmd\nu_\Lcal}{2\pi^{|L|}} \frac{(-\rmi)^{N-1}}{\prod_{n=1}^{N-1} (\Omega^{\overline{P(\Ncal)}}_n - \rmi\eta_n) } \mathcal{D}^{P(\Ncal)}(\nu_\Lcal).
\end{split} \end{equation}
Setting $N^P_- = 0$ leads to a similar expression for an anti-time ordered diagram. These are the types of expressions used by Jeon \cite{Jeon1993} to derive cutting rules for Matsubara functions.
The terms in the sum in \Eq{eq:retarded_equation} can be evaluated with the help of the following diagrammatic rules. 
\begin{enumerate}
    \item Draw the $v$-lines/vertices at the $N$ times in the order of increasing contour time from bottom to top as in figure a) below.
\begin{equation} \label{eq:retarded_rules}
    a) \bigdiagramcenter{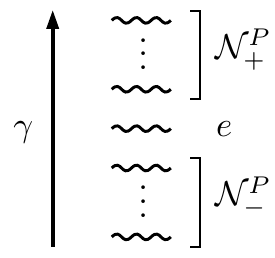} \qquad b) \bigdiagramcenter{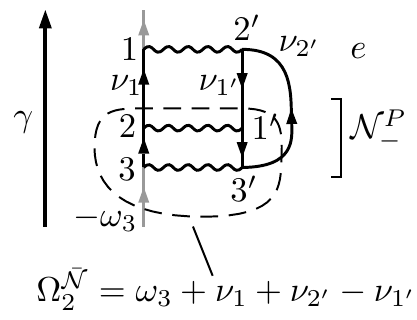}
\end{equation}

    \item Draw the lines between the vertices according to the structure of the diagram. Assign to an ascending $g$-line a greater component $g^>_L(\nu_L)$ and to a descending $g$-line a lesser component $g^<_L(\nu_L)$. Assign to each $v$-line $v_I$. 
    
    \item For each $n \leq N^P_-$ draw a circle around the first $n$ vertices in contour order. For each circle multiply the denominator by the factor $(\Omega - i\eta)$, where $\Omega$ is the total energy flowing outwards through the circle. See figure b) above for an example.
    
    \item For each $n \leq N^P_+$ draw a circle around the last $n$ vertices in contour order. For each circle multiply the denominator by the factor $(\Omega - i\eta)$, where $\Omega$ is the total energy flowing outwards through the circle.
    
    \item Multiply with $C (-1)^{N^P_+} (-\rmi)^{N-1}$, where $C$ is the prefactor for the diagram given by the applicable Feynman rules.
    
    \item Integrate over the frequencies $\nu_j$.
\end{enumerate} 

Let us consider as an example the retarded diagram 
\begin{equation} \label{eq:rules_example}
T^{R(1,2)}(t_1,t_2) = \left[ \bigdiagramcenter{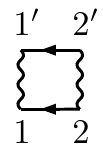} \right]^{R(1,2)}, \qquad \bigdiagramcenter{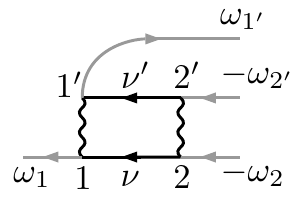}
\end{equation}
that appears as a half-diagram in the retarded expansion for the $T$-matrix approximation. Since time-local interactions connect the vertices pairwise, there are only two contour orderings to consider
since $t_2$ is either earlier or later than $t_1$ on the contour.
Following our rules these give
\begin{align}
    &\bigdiagramcenter{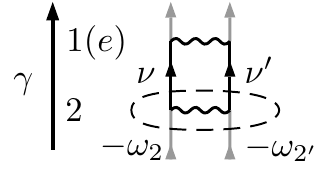} \rightarrow \int \frac{\rmd\nu \rmd\nu'}{(2\pi)^2} \frac{(-1)^0(-\rmi)^1  g^>(\nu) g^>(\nu') v_1 v_2}{\omega_2 + \omega'_2 + \nu + \nu' - \rmi\eta_2} \\
    &\bigdiagramcenter{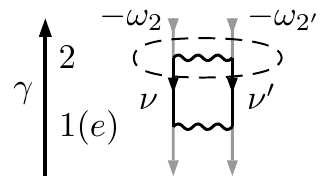} \rightarrow \int \frac{\rmd\nu \rmd\nu'}{(2\pi)^2} \frac{(-1)^1(-\rmi)^1 g^<(\nu) g^<(\nu') v_1 v_2}{\omega_2 + \omega'_2 + \nu + \nu' - \rmi\eta_2} 
\end{align}
where we have substituted $C = 1$ as given by the Feynman rules for our labeling of the external vertices (see \ref{app:prefactors}). Summing over the orderings we therefore have
\begin{equation} \begin{split}
    \mathcal{T}^{R(1,2)}(\bar{\omega}_1,\bar{\omega}_2) 
    &= -2\pi \rmi \delta(\bar{\omega}_1 + \bar{\omega}_2) \int \frac{\rmd\nu \rmd\nu'}{4\pi^2} \frac{g^>(\nu) g^>(\nu')- g^<(\nu) g^<(\nu')}{\bar{\omega}_2 + \nu + \nu' - \rmi\eta_2} v_1 v_2,
\end{split} \end{equation}
where $\bar{\omega}_1 = \omega_1 + \omega_{1'}$ and $\bar{\omega}_2 = \omega_2 + \omega_{2'}$.

Let us now compare the frequency representation for a retarded diagram derived above to the corresponding expression for a Matsubara diagram.
The general expression for an $N$-point Matsubara 
function in frequency space is defined to be
\begin{align}
    \mathcal{D}^M (\omega_\Ncal) &=  \int_{0}^{-\rmi\beta} \rmd z_{\Ncal}\, \rme^{\rmi \omega_\Ncal \cdot z_\Ncal}
    D^M (z_\Ncal) 
    \label{Matsubara_transform1}
 \end{align}
where $\omega_\Ncal=\{ \omega_1,\ldots,\omega_N \}$ is an ordered set of discrete frequencies of the general Matsubara form
$\omega_j = 2\pi \rmi m/\beta$ with $m$ running over
odd integers and $z_\Ncal =\{ z_1,\ldots, z_N \}$ is a set of times $z_j=-\rmi \tau_j$ on the Matsubara branch.
Due to time translation invariance $D^M (-\rmi \tau_\Ncal)= D^M (- \rmi \tau'_\Ncal)$ with
$\tau_j^\prime = \tau_j - \tau_e$, which gives $\tau_e'=0$ on argument position $e$.
A change of integration variable to $\tau_j'$ and use of (anti)-periodic boundary  conditions \cite{Stefanucci2013} then produces an overall Kronecker delta 
multiplied with the factor $-\rmi \beta$ resulting from an imaginary time integral \cite{Baier1994}. This yields
\begin{align}
\mathcal{D}^M (\omega_\Ncal)&= 2\pi \delta_{\omega}\,  \mathscr{D}^{M} (\omega_\Ncal)
\end{align}
in which we defined
$\delta_\omega=\delta_{\omega,0}/(2\pi i/\beta)$ where $\delta_{\omega,0}$ is a Kronecker delta for the sum
$\omega=\omega_1+ \ldots + \omega_N$ of all Matsubara frequencies. 
We further defined the analytic function
\begin{equation}
    \mathscr{D}^{M} (\omega_\Ncal) = (-\rmi)^{N-1}\int_0^\beta \rmd\tau_{\Ncal \setminus e}\, \rme^{\omega_\Ncal \cdot \tau_\Ncal }
    D^M_e (-i \tau_\Ncal)
    \label{Matsubara_transform}
\end{equation}
where $D_e^M$ denotes the Matsubara $N$-point function $D^M (-\rmi \tau_\Ncal)$ in which we put $\tau_e=0$; since $\tau_e=0$ the integral in \Eq{Matsubara_transform}
and therefore $ \mathscr{D}^{M}$ is independent of $\omega_e$.
For the specific diagrammatic form of $D^M$ we assume
an identical expression as in \Eq{structure} where now the contour variables are on 
the vertical Matsubara track.
The Green's function lines can be written as the integrals \cite{Stefanucci2013}
\begin{equation}
    g_L^\lessgtr (\tau,\tau') = \int \frac{\rmd\nu}{2\pi} g_L^\lessgtr (\nu) \rme^{-(\nu-\mu)(\tau-\tau')}. 
\end{equation}

\begin{figure}
    \centering
    \includegraphics[scale=1.0]{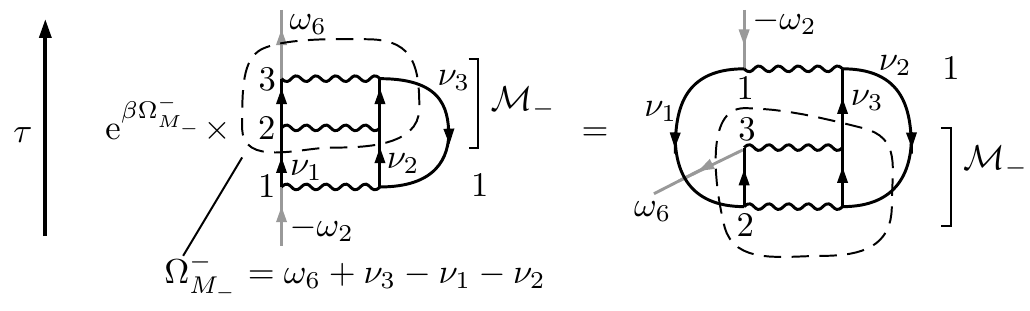} 
    \caption{  Example diagram for \Eq{eq:Mshift}. In this case $\Mcal_{-} = \{ 3,2 \}$, $\Mcal_+ = \emptyset$, $e=1$ and $\Omega_{M_-}^- = \omega_6 + \nu_3 - \nu_2 - \nu_1$.
    Multiplying by $\rme^{\beta \Omega_{M_-}^-}$ converts the left diagram into the right one
    in which the particle-hole lines that enter the circled region enclosing the vertices labeled by $\Mcal_-$ are reversed in direction thereby inducing the cyclic set permutation $(\Mcal_-,\Mcal_+,1) \rightarrow (\Mcal_+,1,\Mcal_-$)}
    \label{fig:baym_rule}
\end{figure}

Inserting these expressions then gives
\begin{align}
    \mathscr{D}^M (\omega_\Ncal) &= 
    \sum_\Mcal \int \frac{\rmd\nu_\Lcal}{(2\pi)^{|L|}} \left[ C \prod_L g_L (\nu_L) \prod_I v_I \right]^{\Mcal e} \Lambda_{\Mcal} (\sigma_\Mcal)
\end{align}
where the sum is over all orderings of the $N-1$ times in $\Mcal=\Ncal \setminus e$
(since we set $z_e=0$ to be the earliest time) and
where we defined
\begin{equation} \label{eq:L_def_2}
\Lambda_{\Mcal} (\sigma_\Mcal) = (-\rmi)^{N-1} \int_0^\beta \rmd\tau_{\Mcal}\, \theta (\tau_\Mcal,0) e^{\sigma_\Mcal \cdot \tau_\Mcal}
\end{equation}
where
\begin{equation}
    \sigma_j =\omega_j + \sum_{L=L_i} \bar{\nu}_L - \sum_{L=L_f} \bar{\nu}_L,
    \label{eq:sigma_sum_M}
\end{equation}
with $\bar{\nu}_L=\nu_L-\mu$.
The latter term can be evaluated using the expression
\begin{align}
    \int_0^\beta \rmd\tau_n \ldots \rmd\tau_2 \,\theta (\tau_n, \ldots, \tau_2,0) \rme^{\sigma_n \tau_n+ \ldots + \sigma_2 \tau_2} 
    &= \sum_{\textrm{div}(\Mcal)} \frac{(-1)^{M_+} \rme^{\beta \Omega_{M_-}^-}}{\prod_{l=1}^{M_-} \Omega_{l}^- \prod_{k=1}^{M_+} \Omega_k^+}
\end{align}
where the sum is over all $n$ divisions $\textrm{div}(\Mcal)$ of the set $\Mcal= \{ n,\ldots,2 \} = \{ \Mcal_-, \Mcal_+ \}$
into two disjoint subsets of size $M_-$ and $M_+$ where we ordered the index corresponding to the latest time to the left. Here $\Omega_l^-$ and $\Omega_{l}^+$ are defined to be the sum over
the first $l$ elements and the last $l$ elements of $\sigma_\Mcal = \{ \sigma_n,\ldots,\sigma_2 \}$ respectively. This expression
can be conveniently derived using Laplace transforms \cite{Bloch1958,Jeon1993} or directly \cite{Baym1963} by repeated integration and rearrangement of the resulting terms.
This finally gives
\begin{equation}
    \mathscr{D}^M(\omega_\Ncal) = \sum_\Mcal \sum_{\textrm{div}(\Mcal)}
    \int \frac{\rmd\nu_\Lcal}{(2\pi)^{|L|}}
    \frac{(-\rmi)^{N-1} (-1)^{M_+} \rme^{\beta \Omega_{M_-}^-}}{\prod_{l=1}^{M_-} \Omega_{l}^- \prod_{k=1}^{M_+} \Omega_k^+} \left[ C \prod_L g_L (\nu_L) \prod_I v_I \right]^{\Mcal e}.
\end{equation}
Now
\begin{equation}
    \rme^{\beta \Omega_{M_-}^-} \, \left[  \prod_L g_L (\nu_L) \prod_I v_I \right]^{\Mcal_- \Mcal_+ e} 
    = \left[  \prod_L g_L (\nu_L) \prod_I v_I \right]^{\Mcal_+ \,e\,\Mcal_-}
    \label{eq:Mshift}
\end{equation}
so the product with the Boltzmann factors causes an effective cyclic permutation of the 
Matsubara times from $(\Mcal_-,\Mcal_+, e)$ to $(\Mcal_+ ,e,\Mcal_-)$.
This is a consequence of the fluctuation-dissipation relations for the greater and lesser
Green's functions: $\rme^{\beta \bar{\nu}} g^< (\nu) = - g^> (\nu)$ \cite{Baym1963}. This is best illustrated
with the example of Figure \ref{fig:baym_rule}, where on the left hand side we consider a particular self-energy diagram for a given time-ordering. In this example the cut lines carry frequencies $\nu_1,\nu_2$ and $\nu_3$ and we have an external vertex with outflowing frequency $\omega_6$.
In this case multiplication with $\rme^{\beta \Omega_{M_-}^-}$ has the effect
\begin{equation}
    \rme^{\beta (\omega_6 + \bar{\nu}_3 - \bar{\nu}_2 - \bar{\nu}_1)} g^> (\nu_1) g^> (\nu_2) g^< (\nu_3) = g^< (\nu_1) g^< (\nu_2) g^> (\nu_3)
\end{equation}
where since $\omega_6$ denotes an odd Matsubara frequency we have $e^{\beta \omega_6 }=-1$.
The particle lines along the cut have turned into hole lines and vice-versa 
leading to a contribution that diagrammatically corresponds to the right hand side of Figure \ref{fig:baym_rule} which amounts to a cyclic permutation of the times of the vertices. 
Since the orderings in $\Mcal$ correspond to all orderings of the times $z_2,\ldots,z_N$ and we also sum over
all ways to insert $e$ into $\Mcal$
we therefore sum over all orderings of $\Ncal=(\Ncal^+ ,e,\Ncal^-)$ and the expression can be rewritten as
\begin{equation}
     \mathscr{D}^M(\omega_\Ncal) = \sum_\Ncal 
    \int \frac{\rmd\nu_\Lcal}{(2\pi)^{|L|}} \frac{ (-\rmi)^{N-1} (-1)^{N_{+}} }{
    \prod_{k=1}^{N_-} \Omega^{\bar{\Ncal}}_k \prod_{l=1}^{N_+} \Omega^\Ncal_l} \left[ C \prod_L g_L (\nu_L) \prod_I v_I \right]^\Ncal.
    \label{eq:DM_final}
\end{equation}
This is a generalisation to multiple external vertices
of an expression derived by Balian and De Dominicis \cite{Balian1960} and Baym and Sessler \cite{Baym1963}
for the frequency representation of the Matsubara self-energy diagrams.
Our expression is of identical form to the one obtained from the Fourier transform of the retarded function (\ref{eq:retarded_equation}). Writing $\omega_i = \bar{\omega_i} + \mu$ in \Eq{eq:sigma_sum_M} we see that this equation becomes identical to \Eq{eq:sigma_sum_R} apart from a shift with $\mu$.
It then follows that the functions $\mathscr{D}^M$ and $\mathscr{D}^R$ are equal if
we replace each $\omega_i$ in $\mathscr{D}^M$ by $\omega_i - \mu -\rmi\eta_i$, i.e.
\begin{equation}
 \mathscr{D}^{R(e,\Ical)} (\omega_\Ncal ) =\mathscr{D}^M(\omega_\Ncal - \mu - \rmi\eta_\Ncal) ,
 \label{analytic_continuation2b}
\end{equation}
where $\eta_i$ are positive infinitesimals (and there is no dependence on $\eta_e$).
This analytic continuation formula is a generalization of \Eq{eq:S_M_S_R_relation} for general diagrams.
Let us see how this relates to the very familiar example of the analytic continuation of the retarded and Matsubara Green's function.
Evaluating the spectral representation of $G^R(t_1,t_2) = G^{R(1,2)}(t_1,t_2)$ using the rules derived above gives for the two contour-orderings
\begin{align}
    \bigdiagramcenter{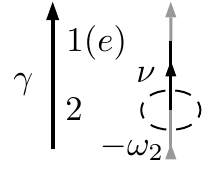} \rightarrow \int \frac{\rmd\nu}{2\pi} \frac{(-1)^0(-\rmi)^1 g^>(\nu)}{\omega_2 + \nu - \rmi\eta_2}, \qquad \bigdiagramcenter{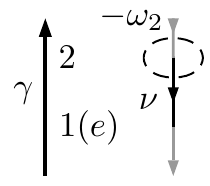} \rightarrow \int \frac{\rmd\nu}{2\pi} \frac{(-1)^1(-\rmi)^1 g^<(\nu)}{\omega_2 + \nu - \rmi\eta_2} 
\end{align}
which give using the spectral function $A(\nu) = \rmi (g^>(\nu) - g^<(\nu))$
\begin{align} \label{eq:GR_spectral}
    G^{R(1,2)} (\omega_1,\omega_2)
    &= -2\pi \delta (\omega_1 + \omega_2) \int \frac{\rmd \nu}{2 \pi} \frac{A (\nu)}{\omega_2 + \nu - \rmi \eta_2} \nonumber \\
    &= 2\pi \delta (\omega_1 + \omega_2) \int \frac{\rmd \nu}{2 \pi} \frac{A (\nu)}{\omega_1 - \nu + \rmi \eta_2} =  2 \pi  \delta (\omega_1 + \omega_2)  G^R (\omega_1).
\end{align}
The rules for evaluating Matsubara functions are exactly the same as those for retarded functions, except that instead of $(\Omega - i\eta)$ the factors appearing in the denominator have the form $(\Omega + \mu)$. These rules therefore give for $G^M$ the spectral representation
\begin{align} \label{eq:GM_spectral}
    G^M (\omega_1,\omega_2)
    &= 2\pi \delta (\omega_1 + \omega_2) \int \frac{\rmd \nu}{2 \pi} \frac{A (\nu)}{\omega_1 - \nu + \mu} =  2 \pi  \delta (\omega_1 + \omega_2)  G^M (\omega_1).
\end{align}
Here $G^R(\omega_1)$ and $G^M(\omega_1)$ are the usual retarded and Matsubara components of the single-particle Green's function in frequency space, and comparing \Eq{eq:GR_spectral} and \Eq{eq:GM_spectral} we find the familiar result
\begin{equation}
    G^R(\omega) = G^M(\omega - \mu + \rmi\eta) .
\end{equation}

An expression similar to \Eq{analytic_continuation2b} was also derived by Baier and Ni{\'e}gawa \cite{Baier1994} based on the assumption that certain mixed imaginary-real time integrals vanish, which is akin to the assumption of a continuous spectrum in combination with the Riemann-Lebesque lemma. They formulate the relation (\ref{analytic_continuation2b}) as
\begin{equation}
 \mathcal{D}^{R(e,\Ical)} (\omega_\Ncal ) = \mathcal{D}^M(\omega_\Ncal - \mu - \rmi\eta_\Ncal) ,
 \label{analytic_continuation2}
\end{equation}
where $\eta_i$ for $i \neq e$ are positive and $\eta_e = - \sum_{i \neq e} \eta_i$.
Our derivation is both more well-defined and more general and has the additional advantage of not requiring the assumption of having a continuous spectrum which we used in deriving the finite temperature cutting rules (Section \ref{sec:time_ordered_cutting}). 
The main result (\ref{analytic_continuation2b}) can now be employed together with the technique of Jeon \cite{Jeon1993} to 
derive the retarded cutting rules directly from the Matsubara ones without the assumption of having a continuous spectrum. 

\printbibliography

\end{document}